\documentclass[sn-basic]{sn-jnl}% Basic Springer Nature Reference Style/Chemistry Reference Style
\jyear{2021}%

\theoremstyle{thmstyleone}%
\newtheorem{theorem}{Theorem}%  meant for continuous numbers
\newtheorem{proposition}[theorem]{Proposition}% 
\newtheorem{lemma}{Lemma}

\theoremstyle{thmstyletwo}%
\newtheorem{example}{Example}%
\newtheorem{remark}{Remark}%

\theoremstyle{thmstylethree}%
\newtheorem{definition}{Definition}%

\newcommand{\bx}{\mathbf{x}}
\newcommand{\by}{\mathbf{y}}
\newcommand{\cX}{{\cal X}}
\catcode`\|=12
\begin{document}

\title[Convergence of MTM-IS]{Convergence Rate of Multiple-Try Metropolis Independent Sampler}
\date{10/28/2022}

\author[1]{\fnm{Xiaodong} \sur{Yang}}\email{yangxiaodong0912@gmail.com}

\author*[1]{\fnm{Jun S.} \sur{Liu}}\email{jliu@stat.harvard.edu}

%\affil[1]{\orgdiv{School of Gifted Young}, \orgname{University of Science and Technology of China}, \orgaddress{\street{Jinzhai Road}, \city{Hefei}, \postcode{230026}, \state{Anhui}, \country{China}}}

\affil[1]{\orgdiv{Department of Statistics}, \orgname{Harvard University}, \orgaddress{\street{Oxford Street}, \city{Cambridge}, \postcode{02138}, \state{Massachusetts}, \country{U.S.}}}


\abstract{The {Multiple-Try Metropolis} method is an interesting extension of the classical {Metropolis-Hastings} algorithm. However, theoretical understanding about its usefulness and convergence behavior is still lacking. We here derive the exact convergence rate for the Multiple-Try Metropolis Independent sampler (MTM-IS) via an explicit eigen analysis. As a by-product, we prove that an naive application of the MTM-IS is less efficient than  using the simpler approach of ``thinned" independent Metropolis-Hastings method at the same computational cost. We further explore more variants and find it possible to design more efficient algorithms by applying MTM to part of the target distribution or creating correlated multiple trials.}




\maketitle
\section{Introduction}\label{sec 1}

\subsection{Fundamental Metropolis-Hastings method}
Markov chain Monte Carlo (MCMC) methods have played important roles in statistical computing and Bayesian inference and have attracted much attention from both theoretical researchers and practitioners. In a nutshell, the set of methods provide general and practical recipes for generating random draws from any given target probability distribution known up to a normalizing constant.
%When facing a complicated probability distribution, Markov chain Monte Carlo methods provide powerful simulation tools to draw samples from this target distribution. 
Specifically, such an algorithm generates a time-homogeneous Markov chain with its stationary distribution being the target one. Under mild assumptions, this chain  converges to the target distribution geometrically \citep{roberts1996geometric,liu1995covariance}. See \cite{liu2008monte} and \cite{brooks2011handbook} for  more comprehensive reviews. The scheme first proposed by \cite{metropolis1953equation} and then generalized by \cite{hastings1970monte}  is arguably the most popular and fundamental construction among all MCMC methods. Let $\pi(\cdot)$ denote the target probability distribution/density function on the state space $\mathcal{X}$. The \textit{Metropolis-Hastings} method constructs a Markov chain $x^{(1)}, x^{(2)},\ldots, $ on $\mathcal{X}$ as follows. At step $t+1$, it proposes a new state $y$ from a user-specified transition function $p(x,y)$, i.e.,  $ y\sim p(x^{(t)},\cdot)$. Then, the next state $x^{(t+1)}$ is equal to $y$ with probability $\rho$ and to $x^{(t)}$ with probability $1-\rho$, where
%the proposal is accepted with probability 
\begin{equation*}
	\rho=\min\left\{1,\frac{\pi(y)p(y,x^{(t)})}{\pi(x^{(t)})p(x^{(t)},y)}\right\}.
\end{equation*}
This design ensures that the generated Markov chain satisfies {\it the detailed balance}  with respect to  $\pi$, which guarantees the chain's reversibility and  convergence under mild conditions.

\subsection{Geometric convergence}
A Markov chain with transition function $A$ is said to be geometrically ergodic if, for $\pi$-almost everywhere $x$, $ \Vert A^n(x,\cdot)-\pi(\cdot)\Vert\leq C(x)r^n$ 
%the following inequality
holds true with  constant $r\in (0,1)$. Here $\Vert\cdot\Vert$ denotes a  distance metric between two probability measures, usually taken as the total variation (TV) distance. Other modes of convergence, such as convergence in $\chi^2$-distance (which implies the convergence in total variation),  have also been investigated \citep{liu1995covariance,liu2008monte}. Establishing this inequality and deriving sharp bounds on the rate $r$ are seen as central tasks in studying MCMC algorithms \citep{tierney1994markov,liu1995covariance,roberts1996geometric}. 

As a generalization of the standard {Metropolis-Hastings} algorithm, the {Multiple-Try Metropolis} (MTM) scheme as formalized in \cite{liu2000multiple} allows one to draw multiple trials at each step and select one according to a specially designed probability distribution.  Although intuitively the MTM scheme  enables one to escape from local optimums more easily, there is little theoretical understanding of the convergence rate of any form of the MTM algorithm, making it a challenging practical concern when deciding whether a MTM approach should be employed for a specific problem. Existing theoretical results on the {Metropolis-Hastings} algorithm clearly cannot be easily extended to the MTM algorithm.
Indeed, getting sharp bounds on the convergence rate of any general-purpose {Metropolis-Hastings} algorithm can be extremely challenging, except for the {Independent Metropolis-Hastings} (IMH) algorithm (which is also called the Metropolised independence sampler  by \cite{liu1996metropolized} and the independence Metropolis chain by \cite{tierney1994markov}). We are therefore tempted to consider whether the IMH's multiple-try version, which we call the \textit{Multiple-Try Metropolis Independent sampler} (MTM-IS), can be tackled theoretically.

\subsection{Convergence rate of Independent Metropolis-Hastings algorithm} 
Geometrical ergodicity is not guaranteed for a general {Metropolis-Hastings} algorithm  unless we impose suitable restrictions \citep{roberts1996geometric}, and exact convergence rates for  {Metropolis-Hastings} algorithms are rare to find \citep{diaconis1998we}.
In practice, geometric ergodicity is often established under the `drift-and-minorization' framework \citep{diaconis2008gibbs}. But this technique usually results in a very conservative bound of the convergence rate, not quite practically useful. 
Because of the very special structure of the  IMH algorithm, explicit eigen-analyses of its transition matrix for the finite-discrete state space case were obtained by  \cite{liu1996metropolized}, which results in the exact convergence rate of  the IMH algorithm  (also a very tight bound on the constant in front of the rate) and offers a comparison with classical rejection sampling and importance sampling. \cite{atchade2007geometric} studies the continuous case by determining the full spectrum of the transition operator of the IMH algorithm. A recent preprint of \cite{wang2020exact} combines previous results and provides a lower bound, hence determining the exact convergence rate. In this paper, we impose similar conditions on the MTM-IS and  study its exact convergence rate.

\subsection{Multiple-Try Metropolis and its variants}\label{sec:MTM1}  
The original idea of  {Multiple-Try Metropolis} (MTM) comes from chemical physicists interested in molecular simulations \citep{frenkel1996understanding}. Its general formulation constructed in \cite{liu2000multiple}
inspires the development of Ensemble MCMC methods by \cite{neal2011mcmc}, connects with particle filtering \citep{martino2014multiple}, and stimulates ideas of parallelizing MCMC \citep{calderhead2014general,yang2018parallelizable}. We refer interested readers to the review of \cite{martino2018review}. Intuitively, the  MTM approach enables one to explore the sample space more broadly, and thus potentially gains efficiency in avoiding being trapped in local modes. The method has been incorporated in some applications such as model selection \citep{pandolfi2010generalization} and Bayes factor estimation \citep{dai2020monte}.

In the context of molecular simulations \citep{frenkel1996understanding}, the multiple-try strategy is often applied to a target distribution in which the state space can be partitioned into two parts: position and orientation, i.e., $\mathbf{x}=(\mathbf{x}^p, \mathbf{x}^o)$. %and correspondingly $\pi(\mathbf{x})\propto \pi(\mathbf{x}^o,\mathbf{x}^p)$. 
For a given $\mathbf{x}^p$, evaluating multiple configurations corresponding to different orientations, $\pi(\bx^p,\bx^{o1}),\ldots, \pi(\bx^p,\bx^{om})$ is not much more expensive than evaluating a single $\pi(\bx^p,\bx^o)$. Thus,  MTM can be quite useful in facilitating an efficient move: we can propose the new configuration by (a) first proposing the  position $\bx^p_{(new)}$; (b) associating with it multiple orientations $\bx^{o1}_{(new)},\ldots,\bx^{om}_{(new)}$; (c) picking one from them properly, and (d) using the MTM rule to do acceptance/rejection. 
%Then, given $\mathbf{x}^p$ we can apply the MTM strategy to sample from $\pi(\mathbf{x}^o|\mathbf{x}^p)$.
%In this case, MTM is useful in that evaluating $\pi(\mathbf{x}^o|\mathbf{x}^p)$ for various $\mathbf{x}^0$ is often much cheaper than dealing with the whole $\pi(\mathbf{x}^0,\mathbf{x}^p)$ altogether.
In addition to this case, MTM is also particularly useful when combined with directional sampling, as in \citep{liu2000multiple,dai2020monte}. Specifically, given a sampling direction $\mathbf{e}$ at position $\mathbf{x}$, multiple trials are drawn simultaneously as $r_1,\ldots,r_m\sim p(r)$ to construct $\mathbf{y}_j=\mathbf{x}+r_j\mathbf{e}$.

Several variants of  the MTM are worth mentioning:
\cite{craiu2007acceleration} propose to use correlated trials to accelerate  MTM and introduces antithetic and stratified sampling to bring correlation; \cite{casarin2013interacting} argue that multiple independent trials from different distributions are worth considering, and connect to interactive sampling algorithms.
Theoretically, \cite{bedard2012scaling} conducts a scaling analysis for {MTM}.
However, to the best of our efforts, we can not find any existing result on the  convergence rate  of an MTM algorithm. 

In this paper, we report the exact convergence rate of the MTM-IS for  general target $\pi(\cdot)$ and proposal $p(\cdot)$. The result is somewhat surprising as it shows that the MTM-IS with $k$ multiple tries is not as efficient as simply repeating the standard IMH algorithm $k$ times, thus suggesting that the we may want to design the $k$ multiple proposals to be ``over-dispersed"  (e.g., negatively correlated) in order to take advantage of the MTM structure. Another useful scenario, as  discussed previously and detailed in Section~\ref{sec:partialMTM}, is to help proposing a better configuration for a general Metropolis-Hastings algorithm by orienting part of the proposal better via MTM.

The rest of the article is organized as follows. Section \ref{sec 2} carries out an eigenvalue analysis of MTM-IS; Section \ref{sec 3} specifies the exact convergence rate of MTM-IS under the total variation distance and offers an inequality to compare  MTM-IS with its corresponding ``thinned" IMH algorithm (i.e., taking one draw from every $k$ iterations of the sampler); Section~\ref{sec:numerical} provides some empirical results for multivariate Gaussian and Gaussian mixtures;
Section~\ref{sec 4} discusses several variants and extensions of MTM; and Section \ref{sec 5} concludes the article with a short remark.

\section{Eigen-analysis of Multiple-Try Metropolis Independent sampler}\label{sec 2}

\subsection{Notations}
Throughout the article, we use $\mathcal{X}$ to denote the state space, which can be either discrete or continuous. Notations $\pi(x)$  and $p(x, y)$ represent the target and proposal distributions, respectively, with $x,y\in \mathcal{X}$. 
If proposal distribution is independent of the current state $x$, we  write it as $p(y)$. The  actual transition function/probability/density  of the MCMC algorithm is denoted by $A(x,y)$.
%from the current state $x\in\mathcal{X}$ to a new state $y$ . 
A collection of multiple trials of size $k$ is written as $\by=(y_1,\ldots,y_k)$. We consider the total variation distance for any two (signed) measures $P$ and $Q$, which is defined as $\Vert P-Q\Vert_{TV} =\sup_{A\in{\cal F}} |P(A)-Q(A)|$, where ${\cal F}$ denotes the $\sigma$-field common to $P$ and $Q$ (e.g., the Borel $\sigma$-field for most common uses).  In Section \ref{sec 4}, we slightly abuse the notation by letting $p(x,\by)$ be the proposal distribution for $x\in\mathcal{X}$ and $\by=(y_1,\ldots,y_k)\in\mathcal{X}^k$, as we would consider multiple {correlated} trials in this section. Besides, we write $p(x,\by_{(-j)}\mid y_j)$ as the conditional distribution of $\by_{(-j)}\equiv (y_1,\ldots,y_{j-1},y_{j+1},\ldots,y_k)$ given $y_j$ and $x$. Lastly, $p_j(x,y_j)=\int p(x, \by ) \mathrm{d} \by_{-j}$ denotes the  conditional marginal distribution of $y_j$ given $x$.

\subsection{Description of the algorithms}
The general framework of the MTM %{Multiple-Try Metropolis} 
as formulated in \cite{liu2000multiple} is summarized in Algorithm \ref{MTM-general}. Let the current state be $x$,  and let the number of multiple tries be $k$.  With a proposal transition function $p(x,y)$ that defines the conditional distribution of $y$, we define the  \textit{generalized importance weight} as
\begin{equation}\label{importance weight}
w(y\vert x)=\frac{\pi(y)}{p(x, y)}\lambda(x,y)
\end{equation}
where $\lambda$ is a symmetric non-negative function (i.e., $\lambda(x,y)=\lambda(y,x)\geq 0$, $\forall x, y$). Thus, the acceptance/rejection ratio in a general MH algorithm is just the ratio of the generalized importance weights.

\begin{algorithm}
\caption{Multiple-Try Metropolis: the current state is $x$.}\label{MTM-general}
\begin{algorithmic}[1]
\State Generate multiple trials $y_1,\ldots,y_k$ independently from $p(x,\cdot)$; 
compute their respective weights $w(y_j\mid x)$ as defined in (\ref{importance weight}) for $j=1,\ldots,k$.
\State Select index $J$ with probability proportional to $w(y_j\mid x)$ and define $y=y_J$.
\State Draw $x_1^\ast,x_2^\ast,\ldots,x_{k-1}^\ast$ independently from $p(y,\cdot)$. And set $x_{k}^\ast=x$.
\State Accept $y$ with the ratio $\rho=\min\left\{1,\frac{\sum_{j}w(y_j\mid x)}{\sum_{j}w(x_j^\ast\mid y)}\right\}$.
\end{algorithmic}
\end{algorithm}

Here, $x_1^\ast,x_2^\ast,\ldots,x_{k-1}^\ast$ are called \textit{balancing trials}, which are drawn to guarantee the detailed balance.  \cite{liu2000multiple} also extend the MTM for generating non-independent multiple trials such as semi-deterministic ones along a direction. If we choose $p(x,y)=p(y)$,   we can modify this algorithm to avoid drawing   additional balancing trials as the algorithm is still valid if we simply replace the $x_j^\ast$ by $y_j$ in computing $\rho$. This modified version is summarized in Algorithm \ref{MTM-IS($k$)} and named the MTM-IS($k$). In this case, we further select $\lambda(x,y)\equiv1$ then the generalized importance weight \eqref{importance weight} turns out to be $w(y\mid x)=\pi(y)/p(y)$, coinciding with the standard notation of importance ratio. In order to simplify the notations, we could write
\begin{equation}
w(y)=\pi(y)/p(y).\label{ratio}
\end{equation}

\begin{algorithm}
	\caption{MTM-IS: the current state is $x$.}\label{MTM-IS($k$)}
	\begin{algorithmic}[1]
		\State Draw multiple trials $y_1,\ldots,y_k$ independently from $p(\cdot)$; compute $w(y_j)$ as defined in \eqref{ratio} for $j=1,\ldots,k$.
		\State Select index $J$ with probability proportional to $w(y_j)$ and define $y=y_J$.
		\State Compute $W=\sum_{j=1}^{k}w(y_j)$.
		\State Accept $y$ with the following probability $\rho=\min\left\{1,\dfrac{W}{W-w(y)+w(x)}\right\}$.
	\end{algorithmic}
\end{algorithm}

%We then add some more remarks on \eqref{ratio}. 
In theory, we assume that $\pi$ is absolutely continuous with respect to $p$, so that this importance weight can be interpreted as the Radon-Nikodym derivative. In practice, one should always choose $p$ so that its support covers that of $\pi$ for the algorithm to work well. The main result of this section is stated in Theorem \ref{spectrum thm}, which can be viewed as a generalization of the results in \cite{liu1996metropolized} and \cite{atchade2007geometric} and provides the exact convergence rate of MTM-IS.

\subsection{Transition distribution decomposition}
%In this subsection, we derive an explicit formula for the transition distribution with current state $x$.

\begin{theorem}\label{thm:H}
	The transition distribution of  MTM-IS %\textit{Multiple-Try Metropolis Independent sampler} 
 can be decomposed as
	\begin{equation}\label{transition distribution}
	A(x,\mathrm{d}y)=R(x)\delta_x(\mathrm{d}y)+\min\{H_k[w(x)],H_k[w(y)]\}\pi(y)\mathrm{d}y,
	\end{equation}
	where $H_k$ is defined  as
	\begin{equation}\label{H function}
	H_k(z)=k\underbrace{\int\ldots\int}_{k-1} \frac{1}{z+\sum_{i=1}^{k-1}w(y_i)}\prod_{i=1}^{k-1}p(y_i)\mathrm{d}y_i,
	\end{equation}
	and $R(x)=1-\int_\mathcal{X}\min\left\{H_k[w(x)],H_k[w(y)]\right\}\pi(y)\mathrm{d}y\in[0,1]$ denotes the rejection probability when the current state is $x\in\mathcal{X}$. In particular, $H_k(z)$ is a strictly decreasing function in $z$. For $k=1$, $H_k$ degenerates to $H_1(z)=z^{-1}$.
\end{theorem}
\begin{proof}
	Let $x\notin B\subset\mathcal{X}$ be measurable, the probability of proposing an element in $B$ and accepting it is
	\begin{align*}
	A(x,B)
	=&\mathbb{P}\left[\bigcup_{j=1}^{k}\left\{\left(y_j\in B\right)\cap\left(J=j\right)\cap\left(y_J\text{ gets accepted}\right)\right\}\right]\\
	=&k\mathbb{P}\left[\left\{\left(y_{k}\in B\right)\cap\left(J=k\right)\cap\left(y_k\text{ gets accepted}\right)\right\}\right].
	\end{align*}
	The last equality appears irrelevant to $x$, but the importance ratio $w(x)=\pi(x)/p(x)$ matters when deciding whether or not the chosen $y_J$ is accepted. Furthermore,
	\begin{align*}
	&A(x,B)\\
	=&k\underbrace{\int\ldots\int}_{k-1}\int_B \frac{w(y)}{w(y)+\sum_{j=1}^{k-1}w(y_j)}\min\left[1,\frac{w(y)+\sum_{j=1}^{k-1}w(y_j)}{w(x)+\sum_{j=1}^{k-1}w(y_j)}\right]p(y)\mathrm{d}y\prod_{j=1}^{k-1}p(y_j)\mathrm{d}y_j\\
	=&k\underbrace{\int\ldots\int}_{k-1}\int_B \min\left[\frac{w(y)}{w(y)+\sum_{j=1}^{k-1}w(y_j)},\frac{w(y)}{w(x)+\sum_{j=1}^{k-1}w(y_j)}\right]p(y)\mathrm{d}y\prod_{j=1}^{k-1}p(y_j)\mathrm{d}y_j\\
	=&\int_B \min\left\{H_k[w(x)],H_k[w(y)]\right\}\pi(y)\mathrm{d}y,
	\end{align*}
	where $H_k$ is as defined in (\ref{H function}). Thus, the overall rejection probability is
	\begin{equation}\label{eq:reject_prob}
	R(x)=1-\int_\mathcal{X}\min\left\{H_k[w(x)],H_k[w(y)]\right\}\pi(y)\mathrm{d}y,
	\end{equation}
	and the prescribed decomposition (\ref{transition distribution}) is thus proved.
\end{proof}

Let $w^\ast\triangleq\inf\{u>0:\pi(x:w(x)\leq u)=1\}$  be the {\it essential supremum} of $w(x)$ on $\mathcal{X}$ w.r.t. $\pi(\cdot)$ (i.e., $w^\ast$ is the smallest value such that $w(x)\leq w^\ast$ with $\pi$-probability $1$). Since $H_k(w)$ is a monotone decreasing function of $w$ (Theorem~\ref{thm:H}), we have an upper bound $R(x)\leq 1-H_k(w^\ast)$. Furthermore, since
\begin{equation*}
A(x,\mathrm{d}y)=R(x)\delta_x(\mathrm{d}y)+\min\{H_k[w(x)],H_k[w(y)]\}\pi(y)\mathrm{d}y\geq H_k(w^\ast)\pi(y)\mathrm{d}y,
\end{equation*}
we have the following mixture representation of the transition function, convenient for comparing with $\pi$: 
\begin{equation}\label{decomposition}
A(x,\mathrm{d}y)=H(w^\ast)\pi(y)\mathrm{d}y+[1-H(w^\ast)]q_{\text{res}}(x,\mathrm{d}y),
\end{equation}
where $q_{\text{res}}(x,B):=\dfrac{A(x,B)-H(w^\ast)\pi(B)}{1-H(w^\ast)}$.
This representation can be used to facilitate a coupling argument to prove the geometric convergence of the Markov chain (more details in Section \ref{sec 3}).

\subsection{Spectrum of the transition operator}

Now we provide a result to fully characterize the spectrum of the transition operator induced by the MTM-IS algorithm.  A similar result was derived for the IMH algorithm  by \cite{liu1996metropolized} for the discrete state-space case, and then by  \cite{atchade2007geometric} in general. To be concrete, we introduce the following definitions.
\begin{definition}\label{tr}
	Let $A(x,y)$ be the transition function of a 
	%Let $K$ be a transition operator on $L^2(\pi)$ defined by 
	Markov chain with $\pi$ as its invariant distribution. We define its transition operator $K: \ L^2(\pi) \rightarrow L^2(\pi)$ as
	\begin{equation}
	K f(x)= \int f(y) A(x,y) dy.
	\end{equation}
	It  computes the conditional mean and is called the forward operator in \cite{liu1995covariance}.
\end{definition}
\begin{definition}\label{def:spectrum}
	Let $K_0$ be the restriction of $K$ onto $L^2_0(\pi)$, the orthogonal complement of the constant function of $L^2(\pi)$. Then the {\it spectrum} of  $K_0$ is 
	\begin{equation}\label{spectrum}
	\sigma(K_0)\triangleq\{\lambda\in\mathbb{R}:K_0-\lambda I\text{ is non-invertible}\}.
	\end{equation}
\end{definition}
The {\it essential range} of a function $R$ is
\begin{equation*}
\text{ess-ran}(R)\triangleq\{\lambda\in\mathbb{R}:\pi(x:\mid R(x)-\lambda\mid <\epsilon)>0,\forall\epsilon>0\}.
\end{equation*}

\begin{theorem}\label{spectrum thm}
	Let $K$ be the transition operator defined by the \textit{MTM-IS} algorithm, and let $K_0$ be similarly defined as in Definition~\ref{tr}. Then, $\sigma(K_0)\subseteq \text{ess-ran}(R)$, where $R$ is the rejection probability defined in (\ref{eq:reject_prob}). The equality holds if $\forall$ $\alpha\in \text{ess-ran}(R)$, $\pi\{y:\ R(y)=\alpha\}=0$.
\end{theorem}

Since the proof is mostly technical, we defer it to the Appendix.
From \eqref{eq:reject_prob} and Theorem~\ref{thm:H}, it is obvious to see that an upper bound of $R(x)$ is $1-H_k(w^\ast)$.
This implies that there is a gap between $1$ and the upper edge $1-H(w^\ast)$  of the spectrum $\sigma(K_0)$, provided that $w^\ast<\infty$.
For the finite discrete state-space case, $H(w^\ast)=1/w^\ast$, and $1-H(w^\ast)$ is the exact convergence rate of the  chain.
\section{Convergence Rate and Algorithmic Comparison}\label{sec 3}

\subsection{Convergence in $\chi^2$-distance}\label{sec:converge1}
The $\chi^2$-distance between two probability distributions $\pi$ and $p$ is defined as
\begin{equation}\label{eq:chi_sq_dist}
d_\chi ^2 (\pi,p) = \text{var}_\pi[p(x)/\pi(x)].
\end{equation}
Let $p_n(x)= A_n(p_0, x)$ denote the  distribution of  $X_n$, the state of the Markov chain after $n$ steps from initialization $p_0$.
It was shown in \cite{liu1995covariance} that
$d_\chi (\pi,p_n) \leq \Vert K_0^n \Vert_2 d_\chi (\pi,p_0)$, where
$\Vert\cdot\Vert_2$ is $L^2$-norm of the operator $K_0$.
It is easy to show that \citep{liu1995covariance} \begin{equation}\label{eq:spec_r}
\rho=\lim_{n\rightarrow\infty} \Vert K_0^n \Vert_2^{1/n}
\end{equation}
is the spectral radius of $K_0$ \citep{liu1995covariance}, which is equal to the maximum of $\sigma(K_0)$ in absolute value. As shown in Theorem~\ref{spectrum thm}, this is bounded by $1-H(w^\ast)$. Thus, $d_\chi (\pi,p_n) \leq (1-H(w^\ast))^n d_\chi (\pi,p_0)$. 
It also follows from the Cauchy-Schwartz inequality that

\begin{align}
\Vert p_n -\pi \Vert_{L_1} &= \int\frac{\mid p_n(x) -\pi(x)\mid}{\sqrt{\pi(x)}} \sqrt{\pi(x)} dx  \nonumber \\
&\leq \left[ \int \frac{(p_n(x) -\pi(x))^2}{\pi(x)} dx\right]^{1/2} 
%\left[\int \pi(x) dx\right]^{1/2} 
=d_\chi(\pi,p_n) \label{eq:l1l2} \\
&\leq (1-H(w^\ast))^n d_\chi (\pi,p_0). \nonumber
\end{align} 

Thus, the $L_1$ distance between $p_n$ and the target $\pi$, also known as their {\it total variation distribution} and denoted as $\|p_n-\pi\|_{TV}$, decreases geometrically bounded by the same rate.
%F_0 t(X) = \int t(Y) K_0 (X,Y) dY.

\subsection{Maximal total variation distance}

\begin{definition}
%	Following the terminologies in \cite{wang2020exact}, we l
Let the transition function of a Markov chain be $A(\cdot,\cdot)$, with the corresponding  stationary distribution $\pi(\cdot)$. The \textit{maximal total variation distance} between the Markov chain's $n$-step distribution and $\pi$ is 
	\begin{equation}\label{maximal variation distance}
	d(n)=\text{ess}\sup_{x\in\mathcal{X}}\Vert A^n(x,\cdot)-\pi(\cdot)\Vert_{TV}.
	\end{equation}
	Moreover, the quantity
	\begin{equation}\label{eq:vd_r}
	r= \lim\sup\limits_{n\rightarrow\infty}d(n)^\frac{1}{n}
	\end{equation}
	is called the \textit{exact convergence rate} of the Markov chain. 
\end{definition}

Since the total  variation distance is equivalent to the $L_1$ distance $\Vert p-\pi\Vert_{TV}=2\Vert p-\pi\Vert_{L^1}$ between two probability measures $\pi$ and $p$, it is easy to see from 
definition of \eqref{eq:spec_r}  and equation \eqref{eq:l1l2} that rate $r\leq \rho$. In the following, we use another a coupling argument to validate this upper bound $r$.
Moreover, we will also show that for the transition kernel defined by  Algorithm~\ref{MTM-IS($k$)}, inequality $r\geq\rho$ also holds.
We need the following lemmas to prove our results.
%from \cite{levin2017markov} and \cite{wang2020exact}, respectively.

\begin{lemma}[Coupling]\label{lem:coupling} \citep{levin2017markov} 
	Suppose $(\Psi_t,\widetilde{\Psi}_t)_{t=0}^\infty$ are a pair of Markov chains with the same transition rule satisfying:
	(i) If $\Psi_i=\widetilde{\Psi}_i$ for some $i$, then for any $j\geq i$, $\Psi_j=\widetilde{\Psi}_j$; and
	(ii)  $\widetilde{\Psi}_0\sim\pi$.
	Then, for $\tau=\min\{n:\Psi_n=\widetilde{\Psi}_n\}$, we have a bound
	\begin{equation*}
	\Vert A^n(x,\cdot)-\pi(\cdot)\Vert_{TV}\leq\mathbb{P}(\tau\geq n).
	\end{equation*}
\end{lemma}

\begin{lemma}[Lower bound]\label{lem:lower} \citep{wang2020exact}
	Let  $R(x)$ denote the rejection probability (\ref{eq:reject_prob}) given current state $x$. That is, 
	\begin{equation*}
	R(x)=1-\int\min\left\{H[w(x)],H[w(y)]\right\}\pi(y)\mathrm{d}y.
	\end{equation*}
	Then, we have a lower bound
	\begin{equation*}
	\Vert A^n(x,\cdot)-\pi(\cdot)\Vert_{TV}\geq[R(x)]^n.
	\end{equation*}
\end{lemma}

\begin{theorem}\label{rate theorem}
	Consider the MTM-IS  defined in Algorithm \ref{MTM-IS($k$)} and let $w^\ast<\infty$ be the essential supremum of $w(x)=\pi(x)/p(x)$. Then, the maximal total variation distance of the algorithm to its target distribution $\pi$ is 
	\begin{equation*}
	d(n)=[1-H_k(w^\ast)]^n.
	\end{equation*}
	Thus, the exact convergence rate of the MTM-IS is $1-H_k(w^\ast)$.
\end{theorem}

\begin{proof}
	We will establish that   upper and  lower bounds of $d(n)$ are equal in the limit.
	%\begin{itemize}
	\vspace {2mm}

	\textbf{Upper Bound.} 
	%We can obtain an upper bound for the total variation distance for a Markov chain to reach its  target distribution by a coupling argument \citep{levin2017markov}.
	An upper bound can be obtained by using the coupling idea of Lemma~\ref{lem:coupling}.
	Consider two Markov chains $\{x_t\}$ and $\{\tilde{x}_t\}$ defined by MTM-IS.	Because of the the decomposition \eqref{decomposition},
 we can interpret the actual transition measure $A(x,\cdot)$ as a mixture of $\pi(\cdot)$ and $q_{\text{res}}(x,\cdot)$, and
%	Given the  decomposition (\ref{decomposition}), we 
define the following coupling rule for the two chains. First, we let $x_0=x$ (for some arbitrary $x\in\mathcal{X}$) and assume that $\tilde{x}_0\sim\pi(\cdot)$ as the initialization of these two chains. %Then we describe the joint evolution rules of these two chains. 
Then, suppose that  the two chains are at $x_t$ and $\tilde{x}_t$, respectively, at time $t$. If $x_t=\tilde{x}_t$, then sample $x_{t+1}$ from $A(x_t,\cdot)$ and set $\widetilde{x}_{t+1}=x_{t+1}$. Thus, their future paths coalesce into one. If $x_t\neq \tilde{x}_t$, we draw   $z\sim \text{Bernoulli}(H(w^\ast))$ and sample $x\sim \pi(\cdot)$. We set $x_{t+1}=\widetilde{x}_{t+1}=x$ if $z=1$. Otherwise, we sample $x_{t+1}\sim q_{\text{res}}(x_t,\cdot)$ and $\widetilde{x}_{t+1}\sim q_{\text{res}}(\tilde{x}_t,\cdot)$, independently. 
	
	Our constructions of $\{x_t\}$ and $\{\tilde{x}_t\}$ have the following properties: (i) marginally these two chains both evolve according to $A(\cdot,\cdot)$; (ii) the distribution of $x_t$ is exactly $A^t(x,\cdot)$ and the distribution of $\tilde{x}_t$ is $\pi(\cdot)$; (iii) once $x_t=\tilde{x}_t$ for some $t$, the two chains coalesce into one afterwards.  
	Applying Lemma~\ref{lem:coupling}, we have
	\begin{equation}\label{upper bound}
	\Vert A^n(x,\cdot)-\pi(\cdot)\Vert_{TV}\leq\mathbb{P}(\tau\geq n)\leq[1-H(w^\ast)]^n.
	\end{equation}
	Taking the supremum over $x\in\mathcal{X}$ we have $d(n)\leq[1-H(w^\ast)]^n$.
	
	\vspace{3mm}	
	
	\textbf{Lower Bound:} For the lower bound, we consider the worst case as demonstrated in the proof of Lemma~\ref{lem:lower} in \cite{wang2020exact}.
	In particular, if we can find some $x^\ast$ such that $w(x^\ast)=w^\ast$, then the proof is over; but sometimes this is not achievable, in which case we take advantage of the continuity and monotonicity of $H_k$. For any $\epsilon>0$,   there exists $\delta>0$ such that $H(w)<H(w^\ast)+\epsilon$ once $w^\ast-\delta<w\leq w^\ast$. By the definition of essential supremum, we can always find some $x_\delta\in\mathcal{X}$ such that $w^\ast-\delta<w(x_\delta)\leq w^\ast$, thus
	\begin{equation*}
	d(n)\geq \Vert A^n(x_\delta,\cdot)-\pi(\cdot)\Vert_{TV}\geq R(x_\delta)^n\geq[1-H(w^\ast)-\epsilon]^n,
	\end{equation*}
	since we know from (\ref{eq:reject_prob}) that
	\begin{align*}
	R(x_\delta) 
	%= 1-\int_\mathcal{X}\min\left\{H_k[w(x_\delta)],H_k[w(y)]\right\}\pi(\mathrm{d}y) 
	\geq 1-\int_\mathcal{X}H_k[w(x_\delta)]\pi(\mathrm{d}y)\geq1-H(w^\ast)-\epsilon.
	\end{align*}
	Letting $\epsilon\rightarrow0$, we derive the final result.
	%\end{itemize}
\end{proof}

\subsection{Comparison with the  IMH sampler}
Since one iteration of  MTM-IS is computationally as expensive as $k$-iterations of the IMH  algorithm, we are interested in knowing  which one has a better convergence rate.
We denote the MTM-IS algorithm with $k$ trials as MTM-IS($k$) to emphasize the role of $k$. Correspondingly,
we denote the $k$-fold thinned  IMH algorithm IMH($k$) (i.e., collecting 1 draw after every $k$ steps of the standard IMH).
Note, however, that a clear advantage of MTM-IS($k$) over IMH($k$) is that the former is straightforward to parallelise  as suggested in \cite{calderhead2014general}, which can considerably speed up the algorithm  in practice.

Previously, we obtain the exact convergence rate of MTM-IS($k$) as $1-H_k(w^\ast)$. We rewrite (\ref{H function}) as an expectation form  to gain some insights:
\begin{equation*}
H_k(z)=k\underbrace{\int\ldots\int}_{k-1} \frac{1}{z+\sum_{i=1}^{k-1}w(y_i)}\prod_{i=1}^{k-1}p(y_i)\mathrm{d}y_i=\mathbb{E}_p\left[\frac{k}{z +\sum_{i=1}^{k-1}w(X_i)}\right],
\end{equation*}
where $X_1,\ldots,X_{k-1}$ are independent samples from $p(\cdot)$.
Setting $k=1$, the formula reduces to $H_1(z)=z^{-1}$, which gives rise to the exact convergence rate $1-1/{w^\ast}$ of  the IMH algorithm as shown in \cite{liu1996metropolized} and \cite{atchade2007geometric}. 
The convergence rate of IMH$(k)$ 
%(i.e., every $k$ repeats of the {Independent Metropolis-Hastings} algorithm is counted as one step) 
is then exactly $(1-1/w^\ast)^k$. We have the following main result, whose proof is deferred to the Appendix.

\begin{theorem}\label{comparison thm}
	With the same notations as in Theorem~\ref{rate theorem}, we have
	\begin{equation}\label{comparison inequality}
	1-H_k(w^\ast)=1-\mathbb{E}_p\left[\frac{k}{w^\ast+\sum_{i=1}^{k-1}w(X_i)}\right]\geq\left(1-\frac{1}{w^\ast}\right)^k
	\end{equation}
	for any $k\geq1$, where all $X_i$'s are taken independently from $p(\cdot)$. Thus, MTM-IS($k$) is no more efficient than IMH$(k)$ although the two algorithms are of similar computational cost.
\end{theorem}
This theorem provides the first theoretical guidance on the use of MTM methods. It implies that in this rather simple MTM-IS framework,
%(it is more efficient than a general \textit{Multiple-Try Metropolis} as no balancing proposals are needed) 
multiple independent proposals are not helpful in improving the the mixing of the algorithm. It is not surprising that IMH is preferable when the target distribution is ``easy''  -- after all, the IMH is perfect if the proposal matches the target exactly and having multiple trials is simply a waste. It is  surprising to us, though, that such a preference holds universally.

We speculate that $k$ independent multiple proposals in a general MTM framework are also not more efficient than the corresponding $k$-fold thinned MCMC algorithm. It therefore casts a doubt on the utility of MTM. Our numerical experiences in the past suggest that the MTM strategy is most helpful in jumping among multiple modes  of the target distribution \citep{liu2000multiple,dai2020monte}.
Also as demonstrated in the molecular simulation literature \citep{frenkel1996understanding}, a form of {\it partial} MTM is very useful in building part of the proposal and will be examined in more detail in Section~\ref{sec:partialMTM}. 
More general correlated multiple proposals may also help \citep{craiu2007acceleration} and will be discussed in Sections~\ref{sec:correlatedMT} and \ref{sec:framework}. 
%A third scenario where MTM helps is for molecular simulations mentioned in Section~\ref{sec:partialMTM}, in which MTM is used to build part of a proposed configuration.

\section{Numerical Illustrations}\label{sec:numerical}
We illustrate  the discrepancy between convergence rates of MTM-IS($k$) and IMH$(k)$ numerically. As expected, if the proposal $p$ is already very close to target $\pi$, IMH$(k)$ is significantly better than MTM-IS($k$).
%Otherwise their differences are very minimal. %$w^\ast$ would have been reasonably small. 
The  performance difference of the two algorithms becomes quite minimal if the proposal distribution differs from the target one considerably, i.e., when $w^\ast$  is large. 
%Here we provide three numerical illustrations. The 
In these examples, the explicit convergence rate formula for MTM-IS($k$)  is still complicated, so we use Monte Carlo to approximate the expectation in \eqref{comparison inequality}.

\subsection{Univariate examples}
The first two examples  were previously used in  \cite{liu1996metropolized} to compare the IMH algorithm with importance sampling and rejection sampling and are reexamined here. The third example is a continuous case with an unbounded domain.

\begin{example}\label{exam:discrete}
Let the state space be $\mathcal{X}=\{1,\ldots,m\}$, $p(i)=1/m$ and $\pi(i)=(2m+1-2i)/m^2,p(i)=1/m$. In this case, $w^\ast=2-1/m$ is close to $2$, leading to an approximate convergence rate $0.5$ for the IMH algorithm. Figure \ref{fig:rate_monotune} displays the convergence rates of MTM-IS($k$) and IMH$(k)$ with $m=1000$ and $k$ ranging from $1$ to $10$ computed from $50000$ independent uniform Monte Carlo samples.
\begin{figure}[h]
	\centering
	\includegraphics[width=0.5\linewidth, height=2.1in]{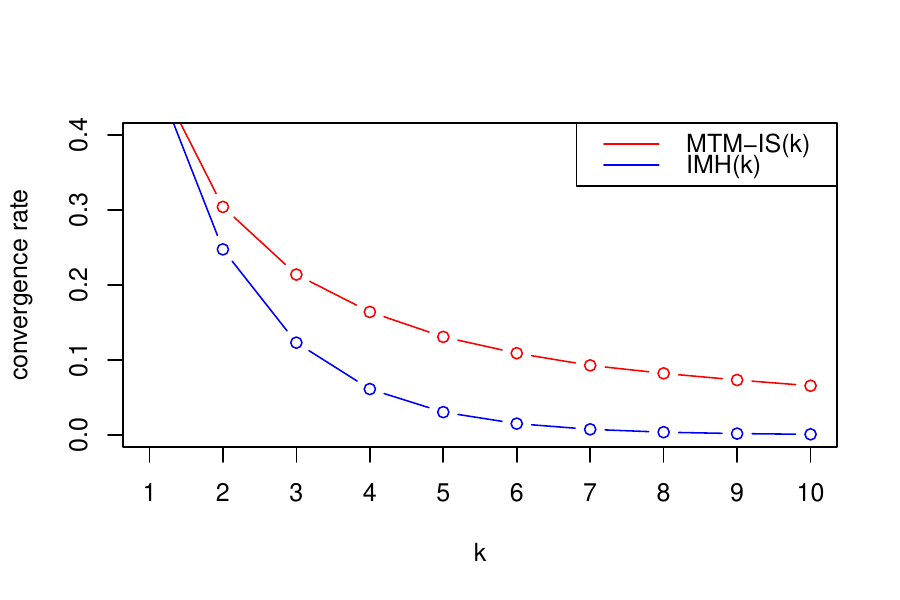}
\caption{Convergence rates  for Example~\ref{exam:discrete} with a finite discrete target distribution.}
	\label{fig:rate_monotune}
\end{figure}
\end{example}

\begin{example}\label{exam:binom}
We consider the case where the target distribution is binomial $\text{Bin}(m,\theta)$, and $p(x)=1/(m+1)$ is uniform. Then
\begin{equation*}
w(x)=(m+1)\frac{m!}{x!(m-x)!}\theta^x(1-\theta)^{m-x}.
\end{equation*}
Using the standard normal approximation, we find that
\begin{equation*}
w^\ast\approx\sqrt{\frac{m}{2\pi\theta(1-\theta)}}.
\end{equation*}
Figure \ref{fig:rate_binomial} is computed from $50000$ independent uniform Monte Carlo samples with $m=100$ for two $\theta$ values. 
%and (a) $\theta=0.5$, (b) $\theta=0.05$. Note
We that in the latter case when the distribution is very skewed, the discrepancy between MTM-IS($k$) and IMH$(k)$ is much smaller.

\begin{figure}[htbp]
	\centering
	
	\begin{minipage}[t]{0.45\linewidth}
		\centering
		\includegraphics[width=1\linewidth,height=2.1in]{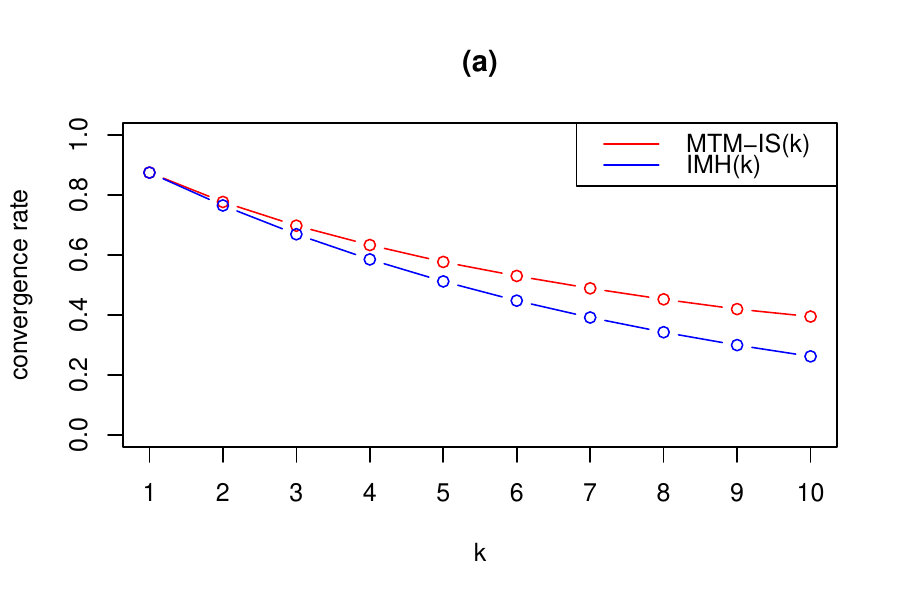}
	\end{minipage}
	\begin{minipage}[t]{0.45\linewidth}
		\centering
		\includegraphics[width=1\linewidth,height=2.1in]{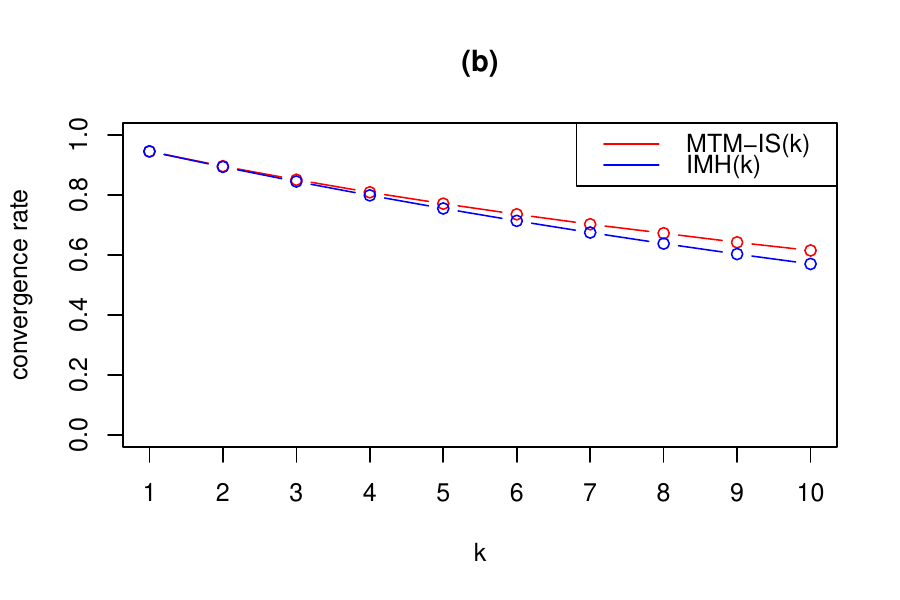}
	\end{minipage}
	\caption{Convergence rates for a binomial target with (a) $\theta$=0.5, and (b) $\theta$=0.05 (Example~\ref{exam:binom}).}
	\label{fig:rate_binomial}
\end{figure}
\end{example}

\begin{example}\label{exam:normal}
We investigate a one-dimensional continuous case with the target being $\mathcal{N}(0,1)$, and the proposal distribution being a scaled t-distribution with $10$ degrees of freedom, $p(x)=ct_{10}(cx)$ with $c\geq1$. For practical uses of both importance sampling and IMH-type algorithms, we strongly recommend to choose a proposal distribution that has a heavier tail than but does not differ too much with the target. 
%A larger $c$ makes the proposal distribution flatter.
 In our case, both t-distribution proposals satisfy the fat-tail requirement. But a larger $c$ leads to a larger discrepancy between the target and the proposal. Figure \ref{fig:rate_continuous} is computed based on $50000$ independent Monte Carlo samples with two choices of $c$, demonstrating that IMH$(k)$ and MTM-IS$(k)$ are nearly indistinguishable if the proposal does not align with the target well. 
 %respect to (a) $c=2$ (b) $c=20$. 

\begin{figure}[htbp]
	\centering
	
	\begin{minipage}[t]{0.45\linewidth}
		\centering
		\includegraphics[width=1\linewidth,height=2.1in]{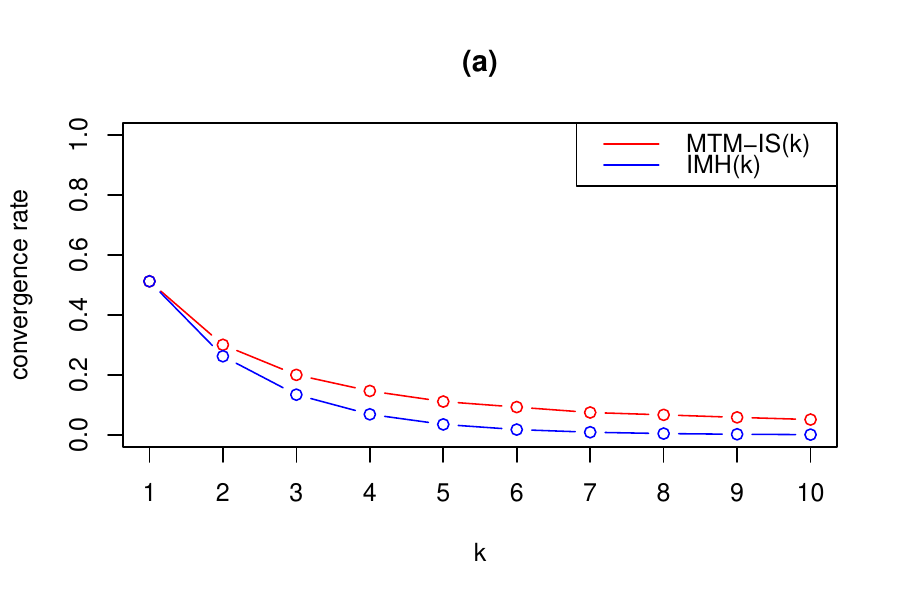}
	\end{minipage}
	\begin{minipage}[t]{0.45\linewidth}
		\centering
		\includegraphics[width=1\linewidth,height=2.1in]{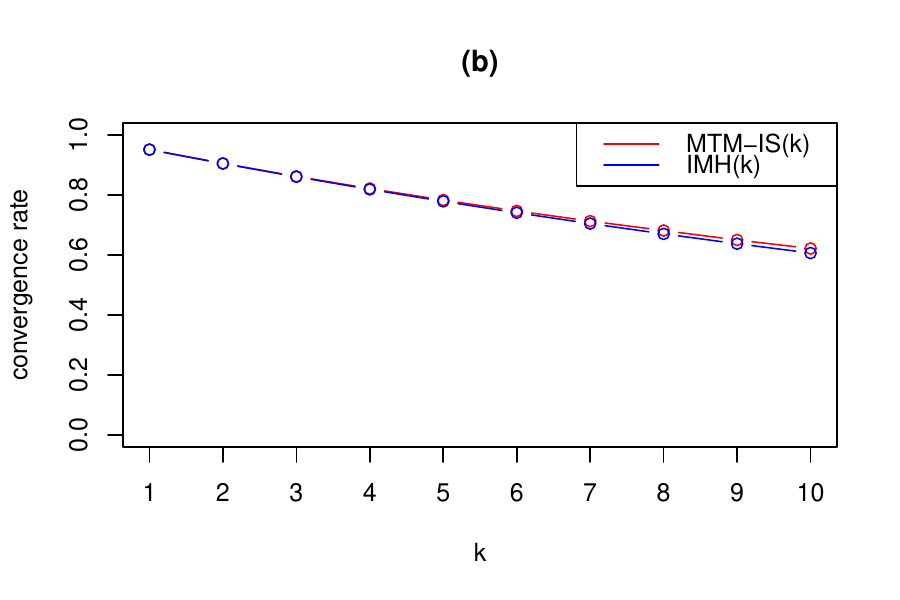}
	\end{minipage}
	\caption{Convergence rates for a standard normal target (Example~\ref{exam:normal})  with the  sampling distribution $p(x)$ being a scaled t-distribution $ct_{10}(cx)$ with (a) $c=2$, (b) $c=20$.}
	\label{fig:rate_continuous}
\end{figure}
\end{example}

\subsection{Multivariate Gaussian and Gaussian mixture}
\begin{figure}[htbp]
	\centering
	
	\begin{minipage}[t]{0.45\linewidth}
		\centering
		\includegraphics[width=1\linewidth,height=2.1in]{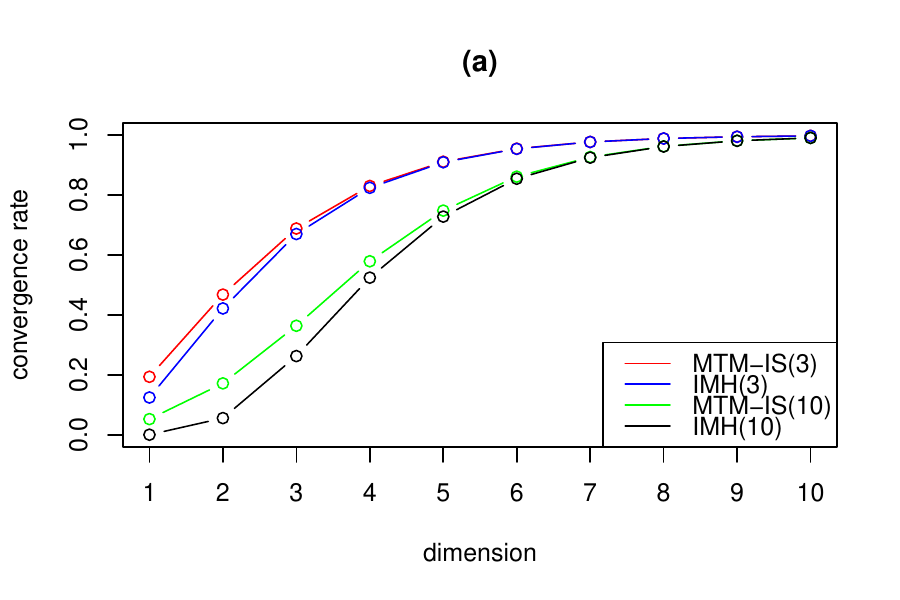}
	\end{minipage}
	\begin{minipage}[t]{0.45\linewidth}
		\centering
		\includegraphics[width=1\linewidth,height=2.1in]{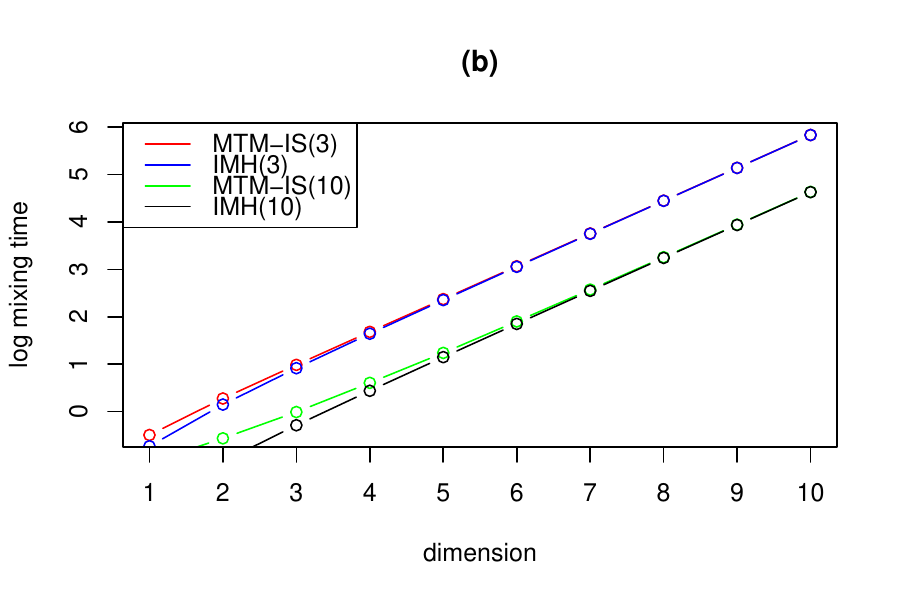}
	\end{minipage}
	\caption{Convergence rates (left) and log-mixing times (right) for the standard multivariate Gaussian target $\pi=\mathcal{N}(0,\mathbb{I}_d)$ with proposal $p=\mathcal{N}(0,4\mathbb{I}_d)$.}
	\label{fig:multi_dim}
\end{figure}

We first use multivariate Gaussian distributions as both the target and proposal to show some practical implications of our result.
Let  $\pi=\mathcal{N}(0,\mathbb{I}_d)$ and $T=\mathcal{N}(\Vec{\mu},\sigma^2\mathbb{I}_d)$. Then we find that the importance weight can be expressed as:
\begin{equation*}
    w(\Vec{x})=\frac{\pi(\Vec{x})}{p(\Vec{x})}=\sigma^d\exp\left[-\frac{1}{2}\left(1-\frac{1}{\sigma^2}\right)\Vert\Vec{x}\Vert^2-\frac{1}{\sigma^2}\langle\Vec{x},\Vec{\mu}\rangle+\frac{1}{2\sigma^2}\Vert\Vec{\mu}\Vert^2\right].
\end{equation*}
Therefore, $w^\ast=\sup w(\Vec{x})<\infty$ if either $\sigma>1$ with an arbitrary $\Vec{\mu}$ or $\sigma=1$ with $\Vec{\mu}=0$. 
%This is quite sensible and consistent with the connection of the IMH and importance sampling \citep{liu1996metropolized}.
When $\sigma>1$, the maximal importance weight $w^\ast\sim\sigma^d$ and thus the mixing time of IMH $\tau_{\text{IMH}}(\delta)=\Omega(w^\ast\log(1/\delta))$ scales exponentially with the dimension $d$. In the same manner, the mixing time of MTM-IS also scales exponentially with $d$, and becomes worse as $\sigma$ increases. Figure \ref{fig:multi_dim} supports that MTM-IS and consecutive IMH have almost the same mixing rates.

\begin{figure}[htbp]
	\centering
	
	\begin{minipage}[t]{0.45\linewidth}
		\centering
		\includegraphics[width=1\linewidth,height=2.1in]{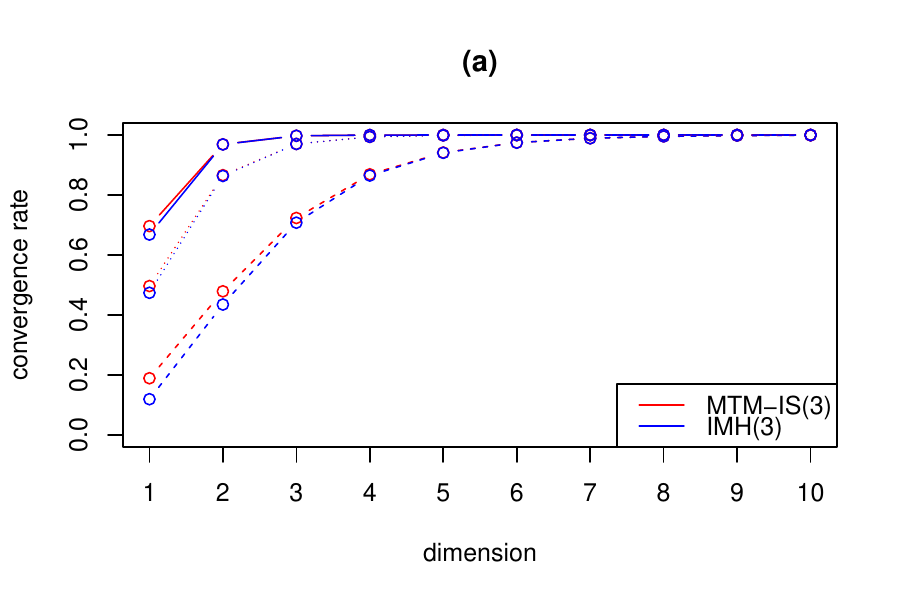}
	\end{minipage}
	\begin{minipage}[t]{0.45\linewidth}
		\centering
		\includegraphics[width=1\linewidth,height=2.1in]{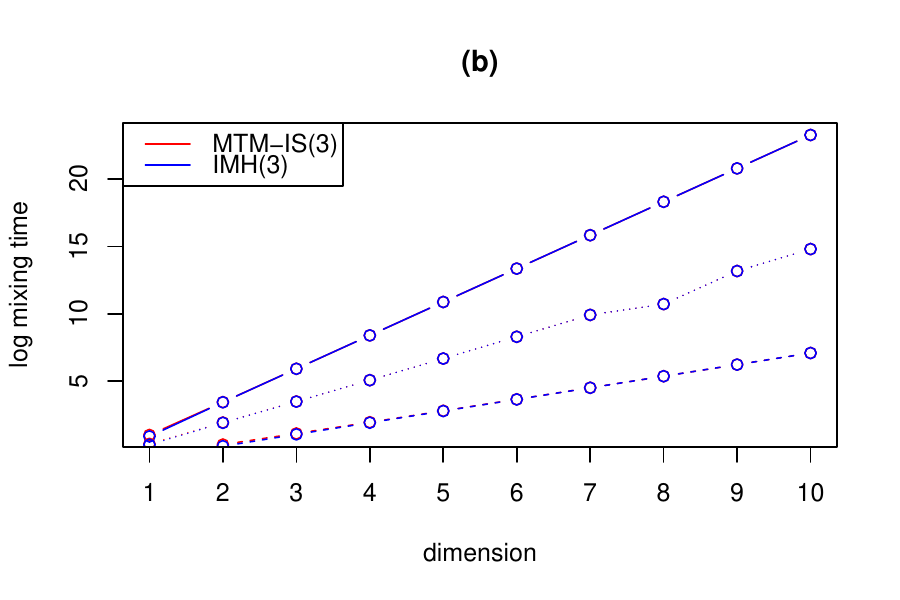}
	\end{minipage}
	\caption{Convergence rates (left) and log mixing times (right) for a multivariate Gaussian mixture target $\pi=\frac{1}{3}\mathcal{N}(0,\mathbb{I}_d)+\frac{2}{3}\mathcal{N}(\Vec{1},\mathbb{I}_d)$ with proposal $p=\mathcal{N}(0,\sigma^2\mathbb{I}_d)$. Solid lines: $\sigma=1.1$; dashed lines: $\sigma=2$; dotted lines: $\sigma=5$. MTM-IS and IMH are nearly indistinguishable.}
	\label{fig:multi_modal}
\end{figure}

Next, we consider  a Gaussian mixture distribution $\pi=\frac{1}{3}\mathcal{N}(0,\mathbb{I}_d)+\frac{2}{3}\mathcal{N}(\Vec{1},\mathbb{I}_d)$, where $\Vec{1}$ is a $d$-dimensional vector filled with all $1$'s. Employing $T=\mathcal{N}(0,\sigma^2\mathbb{I}_d)$, we have the importance weight
\begin{align*}
    w(\Vec{x})=&\frac{1}{3}\sigma^d\exp\left[-\frac{1}{2}\left(1-\frac{1}{\sigma^2}\right)\Vert\Vec{x}\Vert^2\right]\\
    &+\frac{2}{3}\sigma^d\exp\left[-\frac{1}{2}\left(1-\frac{1}{\sigma^2}\right)\Vert\Vec{x}\Vert^2-\frac{1}{\sigma^2}\langle\Vec{x},\Vec{1}\rangle+\frac{d}{2\sigma^2}\right].
\end{align*}
It is easy to see that $w^\ast<\infty$ if and only if $\sigma>1$.
%, implying that still the tail of the distribution matters.
Figure \ref{fig:multi_modal} depicts  theoretical convergence rates and log mixing times for varying dimension and proposal standard deviation $\sigma$. Again the mixing times scale exponentially with dimension $d$. Unlike the single Gaussian case, however, Figure \ref{fig:multi_modal}(b) shows that the slope of log mixing times  is not a monotone function of $\sigma$.

\begin{figure}[htbp]
	\centering
	\begin{minipage}[t]{0.3\linewidth}
		\centering
		\includegraphics[width=1\linewidth,height=2.1in]{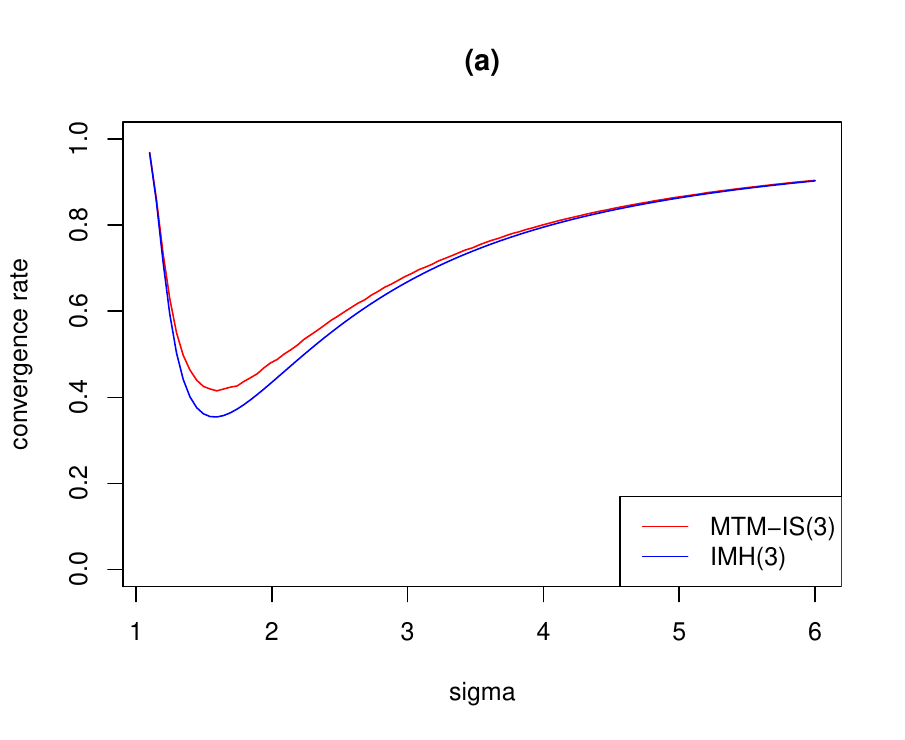}
	\end{minipage}
	\begin{minipage}[t]{0.3\linewidth}
		\centering
		\includegraphics[width=1\linewidth,height=2.1in]{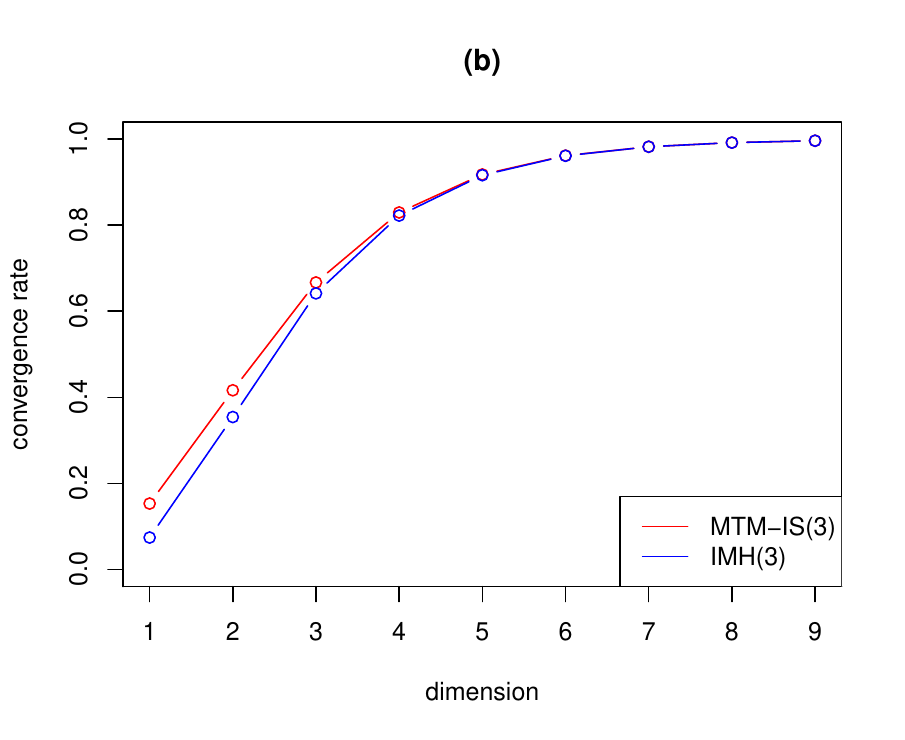}
	\end{minipage}
	\begin{minipage}[t]{0.3\linewidth}
		\centering
		\includegraphics[width=1\linewidth,height=2.1in]{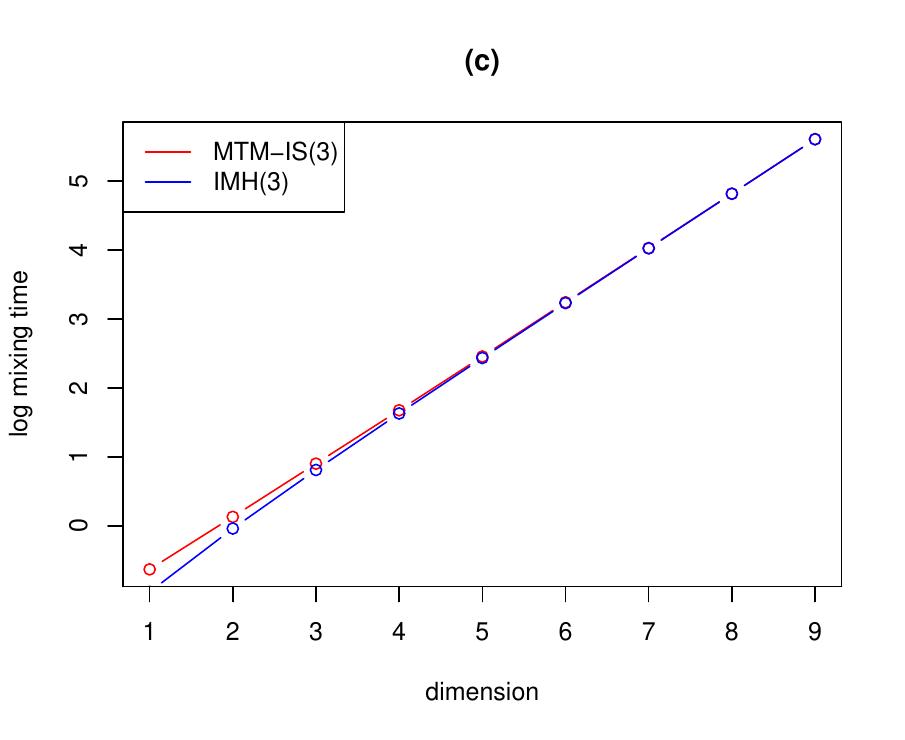}
	\end{minipage}
	\caption{Multidimensional mixture Gaussian: $\pi=\frac{1}{3}\mathcal{N}(0,\mathbb{I}_d)+\frac{2}{3}\mathcal{N}(\Vec{1},\mathbb{I}_d)$ and $T=\mathcal{N}(0,\sigma^2\mathbb{I}_d)$. (a)  convergence rates against $\sigma$ with varying $1.1\leq\sigma\leq6$ when $d$=2; (b) and (c)  plot respectively the convergence rates and log mixing times against the varying dimensions under the optimized $\sigma$.}
	\label{fig:optimal_sigma}
\end{figure}

Figure \ref{fig:optimal_sigma} explores the optimization with $\sigma$. Specifically, Figure \ref{fig:optimal_sigma}(a) plots the convergence rates against varying $\sigma$ when $d=2$, showing that the optimal choice is $\sigma\approx1.594949$. When $d$ grows, the optimal $\sigma$ remains approximately in the range of 1.55$\sim$1.62.  Figure \ref{fig:optimal_sigma}(c)  indicates that the mixing time still scales exponentially with $d$ even if $\sigma$ is optimized. 

\section{Variants of Multiple-Try Metropolis}\label{sec 4}
\subsection{Partial MTM-IS:  an efficient variant}\label{sec:partialMTM}
To reflect how MTM has actually been used in molecular simulations~\citep{frenkel1996understanding}, we assume a partition of the state-space, $\bx=(\bx^a,\bx^b)$, and the corresponding partition of the  target distribution $\pi(\bx)\propto q(\bx^a,\bx^b)=q_a(\bx^a)q_b(\bx^b| \bx^a)$, where $q_b$ may not be normalized. We assume that $q_a(\bx^a)$ is much more expensive to evaluate than $q_b(\bx^b|\bx^a)$.  An important point to note is that we want to move $(\bx^a,\bx^b)$ jointly instead of iterating between conditional draws of $\bx_a|\bx_b$ and $\bx_b|\bx_a$ (for reasons such as the two components may be tightly coupled). We consider the independent proposal: $p(\bx)=p_a(\bx^a)p_b(\bx^b|\bx^a)$.
A {\it Partial MTM-IS} algorithm is as follows: 
\begin{algorithm}[h]
	\caption{PMTM-IS: the current state is $\bx=(\bx^a,\bx^b)$.}\label{MTM-IS-V1($k$)}
	\begin{algorithmic}[1]
		\State Draw $\by^a$ from $p_a(\cdot)$; and draw
		multiple trials $\by_1^b,\ldots,\by_k^b$ independently from $p_b(\cdot\mid \by^a)$;  
		\State Draw  i.i.d. ``balancing trials'' $\bx_1^b,\ldots,\bx_{k-1}^b$  from $p_b(\cdot |\bx^a)$, and let $\bx_k^b =\bx^b$; 
		\State For $j=1,\ldots,k$, compute 
		\[w_j \stackrel{\Delta}{=} %w((\by^a,\by^b_j)) =
		\frac{q_b(\by^b_j\mid \by^a)q_a(\by^a)}{p_b(\by^b_j\mid \by^a)p_a(\by^a)}, \ \ w_j'=\frac{q_b(\bx^b_j\mid \bx^a)q_a(\bx^a)}{p_b(\bx^b_j\mid \bx^a)p_a(\bx^a)},
		\]
  and set $W_y=\sum_{j=1}^{k}w_j$, $W_x=\sum_{j=1}^k w_j'$.
		\State Select index $J$ with probability proportional to $w_j$ and define $\by=(\by^a, \by_J^b)$.
		\State Accept $y$ with probability  $\rho=\min\left\{1,W_y/W_x\right\}$.
	\end{algorithmic}
\end{algorithm}

\begin{remark}[PMTM-IS versus MTM-IS]
Note that, compared with the vanilla MTM-IS
(Algorithm \ref{MTM-IS($k$)}), PMTM-IS needs to draw extra balancing samples. Since we assume that sampling $\bx^b$ and evaluating it are both very cheap, it is still worth doing. In this case, there are no standard IMH or MCMC variants for comparisons.

Typically, one iteration of IMH involves evaluating $q_a/p_a$ twice (respectively on $\bx^a$ and $\by^a$) and evaluating $q_b/p_b$ twice (respectively on $\bx^b|\bx^a$ and $\by^b|\by^a$). In contrast, one iteration of Algorithm \ref{MTM-IS-V1($k$)} consists of evaluating $q_a/p_a$ twice (respectively on $\bx^a$ and $\by^a$) and evaluating $q_b/p_b$ for $2k$ times (respectively on $\bx^b_j|\bx^a$ and $\by^b_j|\by^a$ with $j=1,\ldots,k$). When evaluating $q_b$ is significantly computationally more expensive than $q_a$, Algorithm \ref{MTM-IS-V1($k$)} nearly matches the computational cost of one-step IMH. Under certain reasonable regularity conditions, the following proposition shows that Algorithm \ref{MTM-IS-V1($k$)} provably converges faster.
\end{remark}

\begin{proposition} Let $\bx=(\bx^a,\bx^b)$, and $\pi(\bx)=\pi_a(\bx^a)\pi_b(\bx^b| \bx^a)\propto q_a(\bx_a) q_b(\bx^b|\bx^a)$, where $\pi_a$ and $\pi_b$ are normalized marginal and conditional distributions.
Under the following regularity conditions with proposal $p$ (all parts normalized):
\begin{equation}
    \text{ess}\sup_{\bx^a,\bx^b}\frac{\pi(\bx^a,\bx^b)}{p(\bx^a,\bx^b)}=w^\ast<\infty,
\end{equation}
IMH converges with rate $1-1/w^\ast$. In contrast, the partial MTM-IS (Algorithm \ref{MTM-IS-V1($k$)}) has a convergence rate no slower than $1-1/w^\ast$.
\end{proposition}
\begin{proof}
Noting that $\text{ess}\sup_{\bx}\frac{\pi(\bx^a,\bx^b)}{p(\bx^a,\bx^b)}=w^\ast$, we obtain the convergence rate of IMH as $1-1/w^\ast$ by Theorem \ref{rate theorem}. As for Algorithm \ref{MTM-IS-V1($k$)}, we decompose the transition kernel as
\begin{align*}
    &A((\bx^a,\bx^b),(\by^a,\by^b))\\
    =&k\mathbb{P}\left[\left\{(\by^a\text{ gets proposed})\cap(\by^b_k=\by^b)\cap(J=k)\cap(\text{joint }(\by^a,\by^b)\text{ gets accepted})\right\}\right]\\
    =&k\underbrace{\int_{\mathcal{X}^b}\ldots\int_{\mathcal{X}^b}}_{k-1}\underbrace{\int_{\mathcal{X}^b}\ldots\int_{\mathcal{X}^b}}_{k-1}\frac{p_a(\by^a)p_b(\by^b|\by^a)w_k}{\sum_{j=1}^k w_j}\min\left\{1,\frac{\sum_{j=1}^k w_j}{\sum_{j=1}^k w_j^\prime}\right\}\\
    &\quad\prod_{j=1}^{k-1}p_b(\by^b_j|\by^a)p_b(\bx^b_j|\bx^a)\mathrm{d}\by_j^b\mathrm{d}\bx_j^b.
\end{align*}
Suppose the normalizing constant of $q(\bx^a,\bx^b)$ is $C$, i.e., $\pi(\bx^a,\bx^b)=q(\bx^a,\bx^b)/C$. Then,
\begin{align*}
    &\frac{p_a(\by^a)p_b(\by^b|\by^a)w_k}{\sum_{j=1}^k w_j}\min\left\{1,\frac{\sum_{j=1}^k w_j}{\sum_{j=1}^k w_j^\prime}\right\}\\
    =&\frac{q(\by^a,\by^b)}{\max\left\{\sum_{j=1}^{k}w_j,\sum_{j=1}^{k}w_j^\prime\right\}}=\frac{q(\by^a,\by^b)/C}{\max\left\{\sum_{j=1}^k\frac{q(\by^a,\by^b_j)/C}{p(\by^a,\by^b_j)},\sum_{j=1}^k\frac{q(\bx^a,\bx^b_j)/C}{p(\bx^a,\bx^b_j)}\right\}}\\
    =&\frac{\pi(\by^a,\by^b)}{\max\left\{\sum_{j=1}^k\frac{\pi(\by^a,\by^b_j)}{p(\by^a,\by^b_j)},\sum_{j=1}^k\frac{\pi(\bx^a,\bx^b_j)}{p(\bx^a,\bx^b_j)}\right\}},
\end{align*}
in which $\by_k^b=\by^b$ and $\bx^b_k=\bx^b$. Therefore, it gives rise to
\begin{align*}
    &A((\bx^a,\bx^b),(\by^a,\by^b))\\
    =&k\pi(\by^a,\by^b)\underbrace{\int_{\mathcal{X}^b}\ldots\int_{\mathcal{X}^b}}_{k-1}\underbrace{\int_{\mathcal{X}^b}\ldots\int_{\mathcal{X}^b}}_{k-1}\frac{\prod_{j=1}^{k-1}p_b(\by^b_j|\by^a)p_b(\bx^b_j|\bx^a)\mathrm{d}\by_j^b\mathrm{d}\bx_j^b}{\max\left\{W(\by^a;\by^b_{1:k-1},\by^b),W(\bx^a;\bx^b_{1:k-1},\bx^b)\right\}},
\end{align*}
where $W(\bx^a;\bx^b_{1:k})\triangleq\sum_{j=1}^k\frac{\pi_b(\bx^a,\bx^b_j)}{p_b(\bx^a,\bx^b_j)}$ for any $\bx^a\in\mathcal{X}^a,\bx^b_{1:k}=(\bx^b_1,\ldots,\bx^b_k)\in(\mathcal{X}^b)^k$. By definition, we find
\begin{equation*}
    W(\bx^a;\bx^b_{1:k})=\sum_{j=1}^k\frac{\pi_b(\bx^a,\bx^b_j)}{p_b(\bx^a,\bx^b_j)}\leq kw^\ast.
\end{equation*}
The following inequality immediately follows:
\begin{equation}\label{eq:transition kernel for MTM-IS v1}
    A((\bx^a,\bx^b),(\by^a,\by^b))\geq \frac{\pi(\by^a,\by^b)}{w^\ast}.
\end{equation}
Surprisingly, \eqref{eq:transition kernel for MTM-IS v1} leads to a mixture decomposition like \eqref{transition distribution} and thus is sufficient to construct the upper bound in Theorem \ref{rate theorem} by the coupling argument and Lemma~\ref{lem:coupling}. Therefore, the convergence rate of Algorithm \ref{MTM-IS-V1($k$)} is no larger than $1-1/w^\ast$. However, the arguments for establishing matching lower bounds cannot directly apply due to the extra balancing trials $\bx^b_j,1\leq j\leq k-1$. So the exact convergence rate of Algorithm \ref{MTM-IS-V1($k$)}  remains unknown.
\end{proof}

\subsection{Correlated multiple trials}\label{sec:correlatedMT}
Compared with the original MTM,  the partial MTM-IS differs in that its multiple trials $(\by^a,\by^b_1),\ldots,(\by^a,\by^b_k)$ are correlated due to the state space partitioning. 
%Accompanied by the empirical demonstrations in 
As also demonstrated by \cite{craiu2007acceleration}, we believe that generating correlated
multiple trials is a key in designing efficient MTM algorithms.
%correlations between multiple trials could accelerate the convergence.
%With correlated multiple trials, developing 
Although rigorous theoretical analysis for  a general  correlated MTM design is beyond our reach, we present some theoretical results for two special cases for finite state spaces, which may also be generalization to continuous state-spaces. 
%Although we only study two special cases,  
Implications derived from the analysis apply more generally: good correlated multiple-tries can be obtained with the aid of a deterministic step. 
%Their generalizations to continuous state-spaces are also possible. 
\vspace{3mm}

\noindent
{\bf Stratified  sampling:}
Suppose ${\cal X}$ is a finite state space. We partition it into a few subgroups, $\cX_1, \ldots, \cX_B$ so that $\cX_{i}\cap\cX_j=\emptyset$, $\forall i\neq j$ and $\cup_j \cX_j =\cX$. We begin with a block wise IMH step by sampling from $\{\mathcal{X}_1, \ldots, \mathcal{X}_B\}$ with weight $p(\mathcal{X}_j)$ and accept it with $w(\mathcal{X}_j)=\pi(\mathcal{X}_j)/p(\mathcal{X}_j)$ afterwards. Then, we draw $y$ within the sampled block with probability proportional to $\pi(y)$. It is easy to see that the chain become stationary once it converges at the subgroup level. Thus, the convergence rate of this algorithm is 
\[r_B =1-{w(\cX^\ast)}^{-1},\]
where $\cX^\ast =\arg\max_j w(\cX_j)$. This is not generally better than the IMH$(k)$, which has a convergence rate of
$(1-1/w^\ast )^k> 1- k/w^\ast$. But if the weights $w$'s are very uneven and we can partition the states so that the weights $w(\cX_j)$'s are more balanced, then the stratified IMH can improve upon IMH$(k)$ significantly. We also note that the computation cost of this block-based MTM-IS($k$) algorithm is no worse than IMH$(2)$ (the first step of block sampling is no worse than 1-step IMH; and so is the second step of sampling within a block), much better than IMH$(k)$ when $k$ is large.

\begin{example}[Example 1 continued]
Let $\cX=\{1,\ldots, N\}$, and suppose that the target $\pi(x)\propto x$ and $p(x)\propto 1$. Then, 
the original weights are $w(x) \propto x$ and $w^\ast=\frac{2N}{1+N}\approx 2$. Let $k=2$, then IMH$(2)$ has a rate of $(1-1/w^\ast)^{2}\approx 0.25$, which is quite good.
Assume that $N$ is an even number and  we partition the space as $\cX_j=\{j, N-j+1\}$ for $j=1,\ldots, N/2$. Then $w(\cX_j)\propto 1$, and the resulting MTM-IS(2) converges in one step.
More generally, for an arbitrary distribution $\pi(x)$ and the uniform proposal $p(x)=(2N)^{-1}$, we have $w^\ast=\pi(x^\ast)$ with $x^\ast=\arg\max_x \pi(x)$. Thus, if we can partition the state space so that $\pi(\cX_j)$ are approximately equal for $j=1,\ldots,B$, the algorithm can be much improved. 
\end{example}
\vspace{3mm}

\noindent
{\bf Sampling without replacement:}
Another obvious way of introducing correlations for multiple proposals is to do sampling without replacement. Let $\cX=\{1,\ldots, N\}$. To simplify the discussion, we here focus on the simple random sampling without replacement (SRSWOR, i.e., $p(\by)\propto 1$), although it is possible to extend the method to do sampling without replace with unequal probabilities using one of the schemes in \cite{chen1994weighted}. The algorithm is as follows.

\vspace{2mm}
\noindent

\begin{algorithm}
	\caption{MTM-SRSWOR($k$): Suppose that the current state is at $x$.}\label{alg:MTM-SRSWOR}
	\begin{algorithmic}[1]
		\State Draw  $S=(y_1,\ldots,y_k) \subset(\mathcal{X}\backslash\{x\})$  jointly via SRSWOR.
		\State Select index $J$ with probability proportional to $w(y_j)=(N-1)\pi(y_j)$ and define $y=y_J$.
		\State Accept $y$ with the ratio $\rho=\min\left\{1,\frac{w(y,x)+\sum_{i\neq J}w(y_i,x)}{w(x,y)+\sum_{i\neq J}w_i(y_i,y)}\right\}=\min\left\{1,\frac{\pi(y)+\sum_{j\neq J}\pi(y_j)}{\pi(x)+\sum_{j\neq J}\pi(y_j)}\right\}.$
	\end{algorithmic}
\end{algorithm}

The actual transition probability from $x$ to  $y\neq x$ for this scheme is
\begin{equation}
A(x,y)=\sum_{S^{(k-1)}_y}\frac{1}{\binom{N-1}{k}}\pi(y)\min\left[\frac{1}{\pi(y)+\sum_{i<k}\pi(y_i)},\frac{1}{\pi(x)+\sum_{i<k}\pi(y_i)}\right],
\end{equation}
where $S^{(k-1)}_y\subset\mathcal{X}\backslash\{x,y\}$, $|S^{(k-1)}_y|=k-1$ , and $ y_j \in S^{(k-1)}_y, \forall j<k$. Doing an exact eigenvalue decomposition of matrix $A$ would have brought us a tight bound on the convergence rate. But $A$ does not possess a nice low-rank property as that for the IMH sampler  or the MTM-IS.

For  $S\subset \cX$, we define $\pi(S)=\sum_{x\in\cX}\pi(x)$, $S^\ast =\arg\max_{\{S: \ |S|=k\} } \pi(S)$, and  $x^\ast=\arg\max_x\pi(x)$. We find    the following inequality to hold:
\begin{equation*}
A(x,y)\geq\frac{k\pi(y)}{(N-1)\pi(S^\ast)}, \ \ x\neq y.
\end{equation*}
During each iteration, the chain stays at the current state if and only if the new proposal is rejected since in our construction of Algorithm \ref{alg:MTM-SRSWOR}, the proposal set is not allowed to contain the current state. We observe that $\rho\equiv1$ whenever $x=x_\ast\triangleq\arg\min_x\pi(x)$, leading to $A(x_\ast,x_\ast)=0$. This fact prevents us from using the previous coupling arguments directly. However, as we specify to some circumstances, we could still obtain satisfactory results.

\begin{example}
Choosing $k=2$ and $\mathcal{X}=\{1,\ldots,N\}$,we set
\begin{equation}
\pi_1=1-p,\ \ \pi_2=\cdots=\pi_N=\frac{p}{N-1}, 
\end{equation}
where $0\leq p\leq(N-1)/N$, which guarantees that $x^\ast=1$ and $\{2,\ldots,N\}\in\arg\min_x\pi(x)$. As a result, we know that $A(2,2)=\cdots=A(N,N)=0$. Furthermore, matrix $A$ can be completely determined by the following four quantities:
\begin{align*}
a_1&=A(1,2)=\frac{2\pi_2}{(N-1)(\pi_1+\pi_2)},\\
a_2&=A(1,1)=\frac{\pi_1-\pi_2}{\pi_1+\pi_2},\\
a_3&=A(2,1)=\frac{2\pi_1}{(N-1)(\pi_1+\pi_2)},\\
a_4&=A(2,3)=\frac{(N-3)}{(N-1)(N-2)}+\frac{2\pi_2}{(N-1)(N-2)(\pi_1+\pi_2)}.
\end{align*}
We can then write out $A$ as follows:
\begin{equation}
A=\left[
\begin{matrix}
a_2 & a_1 & a_1 & a_1 & \ldots & a_1\\
a_3 & 0 & a_4 & a_4 & \ldots & a_4\\
a_3 & a_4 & 0 & a_4 & \ldots & a_4\\
a_3 & a_4 & a_4 & 0 & \ldots & a_4\\
\ldots&&&&&\ldots\\
a_3 & a_4 & a_4 & a_4 & \ldots & 0\\
\end{matrix}
\right].
\end{equation}
Now this matrix admits a useful low-rank decoupling: $A=G+ep^T$, where $e=\left[1,\ldots,1\right]^T$, $p=\left[a_3,a_4,\ldots,a_4\right]^T$ and
\begin{equation}
G=\left[
\begin{matrix}
a_2-a_3 & a_1-a_4 & a_1-a_4 & a_1-a_4 & \ldots & a_1-a_4\\
0 & -a_4 & 0 & 0 & \ldots & 0\\
0 & 0 & -a_4 & 0 & \ldots & 0\\
0 & 0 & 0 & -a_4 & \ldots & 0\\
\ldots&&&&&\ldots\\
0 & 0 & 0 & 0 & \ldots & -a_4\\
\end{matrix}
\right].
\end{equation}
Note that $e$ is a common right eigenvector for both $A$ and $A-G$, corresponding to the largest eigenvalue $1$. Since $A-G$ is of rank $1$, the remaining eigenvalues of $A$ and $G$ have to be the same. Hence the eigenvalues for $A$ are $1,a_2-a_3,-a_4,\ldots,-a_4$. This decoupling trick has also been used  in \cite{liu1996metropolized} for  the IMH  algorithm. Given the convergence rate $\left(1-1/(N\pi_1)\right)^2$ of IMH$(2)$, it suffices to show
\begin{equation}
\mid a_2-a_3\mid\leq\left(1-1/(N\pi_1)\right)^2,\ \ a_4\leq\left(1-1/(N\pi_1)\right)^2,
\end{equation}
to prove that MTM-SRSWOR($2$) is faster than IMH$(2)$. Clearly, this holds true  for $p=\dfrac{N-1}{2N}$, which leads to $\pi_1=1/2+1/(2N)$, $\pi_2=1/(2N)$. In this case,
\begin{align*}
a_2-a_3=&1-\frac{4N}{(N+2)(N-1)}<1-\frac{4}{N+1}<\left(1-\frac{2}{N+1}\right)^2=\left(1-\frac{1}{N\pi_1}\right)^2,\\
a_4=&\frac{(N-3)}{(N-1)(N-2)}+\frac{2}{(N-1)(N-2)(N+2)}<\left(1-\frac{1}{N\pi_1}\right)^2.
\end{align*}
\end{example}

We note that designing a suitable parallel construction to do SRSWOR can speed up the algorithm considerably. Furthermore, when proposing multiple trials, we may also choose  not to exclude $x$ from the proposal set. In this case, we  need to modify Algorithm~\ref{alg:MTM-SRSWOR} slightly to  become Algorithm~\ref{alg:MTM-SRSWOR2}. 

\begin{algorithm}
	\caption{MTM-SRSWOR-II($k$): the current state is $x$.}\label{alg:MTM-SRSWOR2}
	\begin{algorithmic}[1]
		\State Draw a subset $S\subset\cX$ of size $k$ at random, denoted as $S=(y_1,\ldots,y_k)$.
		\State Select index $J$ with probability proportional to $w(y_j)=N\pi(y_j)$ and define $y=y_J$.
		\State If $x\not\in S$, accept $y$ with probability $\rho=\min\left\{1,\frac{w(y)+\sum_{j\neq J}w(y_j)}{w(x)+\sum_{j\neq J}w(y_j)}\right\}$. If $x\in S$, accept $y$ with probability $1$.
	\end{algorithmic}
\end{algorithm}

\subsection{Independent non-identical proposals}\label{sec:diff_prop}

Besides introducing correlations between multiple trials, \cite{craiu2007acceleration} also suggests to use different proposals for generating multiple trials in each MTM iteration and provides some supportive empirical evidences.
Here we consider a special case of MTM-IS($k$) in which the multiple trials are  generated from different proposals, i.e., $y_j\sim p_j(\cdot)$ independently for $j=1,\ldots,k$. In this case, we also do not have to draw balancing trials. Defining $w_j(x):=\pi(x)/p_j(x)$, we summarize the procedure in Algorithm~\ref{alg:MTM-IS with different proposal distribution}. 

\begin{algorithm}
	\caption{MTM with independent non-identical proposals with current state $x$.}\label{alg:MTM-IS with different proposal distribution}
	\begin{algorithmic}[1]
		\State Draw multiple trials $y_j\sim p_j(y_j),j=1,\ldots,k$ independently. Then compute $w_j(y_j)=\pi(y_j)/p_j(y_j)$.
		\State Select index $J$ with probability proportional to $w_j(y_j,x)$ and define $y=y_J$.
		\State Accept $y$ with the ratio $\rho=\min\left\{1,\dfrac{w_J(y)+\sum_{i\neq J}w_i(y_i)}{w_J(x)+\sum_{i\neq J}w_i(y_i)}\right\}$.
	\end{algorithmic}
\end{algorithm}
To demonstrate the effect of the \emph{multiple-try} design employed in Algorithm~\ref{alg:MTM-IS with different proposal distribution}, it should be compared with a \emph{sequential $k$-step IMH} sampler. During one iteration, this sampler runs an interior loop of length $k$, within which the $j$-th step proposes an independent proposal from $p_j$ and then accepts/rejects it based on the MH rule as in the ordinary IMH sampler. This sequential IMH sampler has the same computational cost as Algorithm~\ref{alg:MTM-IS with different proposal distribution}.
%Given some more conditions on used proposals $(p_1,\ldots,p_k)$, the
The following theorem provides tight upper bounds for the  convergence rates of the two algorithms, and its proof is deferred to appendixes.
\begin{theorem}\label{final ineq}
    Suppose target $\pi$ is absolutely continuous with respect to every proposal $p_j$.
    Algorithm~\ref{alg:MTM-IS with different proposal distribution} and its corresponding \emph{sequential IMH} sampler are geometrically convergent, with their corresponding respective convergent rates upper bounded by  
    $1-\sum_{j=1}^{k}\mathbb{E}_{p}\left[\frac{1}{w_j^\ast+\sum_{1\leq i\leq k,i\neq j}w_i(X_i)}\right]$ and  $\prod_{j=1}^k \left(1-\frac{1}{w_j^\ast}\right)$, respectively, where $w_j^\ast:=\sup_{x\in\mathcal{X}}w_j(x)$. 
    Furthermore, the following inequality holds,
	\begin{equation}\label{eq:generalized_rate}
	1-\sum_{j=1}^{k}\mathbb{E}_{p}\left[\frac{1}{w_j^\ast+\sum_{1\leq i\leq k,i\neq j}w_i(X_i)}\right]
	\geq\prod_{i=1}^k \left(1-\frac{1}{w_i^\ast}\right),
	\end{equation}
	implying that the upper bound for Algorithm~\ref{alg:MTM-IS with different proposal distribution} is worse than that for the corresponding sequential IMH.
	\end{theorem}

    %and there exists $x^\ast\in\mathcal{X}$ such that for all $j\in\{1,\ldots,k\}$,
\begin{remark}[Tightness of the lower bounds] Suppose $\exists \ x^\ast $ such that
    \begin{equation}\label{eq:condition on MTM with differing proposal}
w_j(x^\ast)=w_j^\ast:=\sup_{x\in\mathcal{X}}w_j(x)=\sup_{x\in\mathcal{X}}\pi(x)/p_j(x)<\infty, \ \text{for all } j,
    \end{equation}
i.e., different proposals  have their importance weight functions $w_j$ to attain their respective supremums at a same point $x^\ast$.
Then, the convergence rates for both aforementioned algorithms
%Algorithm~\ref{alg:MTM-IS with different proposal distribution} and its corresponding sequential IMH 
attain their respective upper bounds.
When $p_1=\cdots=p_k$, condition \eqref{eq:condition on MTM with differing proposal} automatically holds, recovering the convergence rate result 
%of the original MTM-IS (Algorithm~\ref{MTM-IS($k$)}) 
of  Theorem~\ref{rate theorem}. 
However, when there is no such a $x^\ast$  as required by  \eqref{eq:condition on MTM with differing proposal}, the quantities claimed in Theorem~\ref{final ineq} are only upper bounds. It  remains unknown under what other conditions one algorithm can be provably better than the other. Our empirical study shows that their computational efficiencies are almost indistinguishable when the target distribution is ``hard'' relative to the proposals.
\end{remark}

 %   with exact rates $1-\sum_{j=1}^{k}\mathbb{E}_{p}\left[\frac{1}{w_j^\ast+\sum_{1\leq i\leq k,i\neq j}w_i(X_i)}\right]$ and  $\prod_{i=1}^k \left(1-\frac{1}{w_i^\ast}\right)$, respectively. Furthermore, the following inequality holds,
%	\begin{equation}\label{eq:generalized_rate}
%	1-\sum_{j=1}^{k}\mathbb{E}_{p}\left[\frac{1}{w_j^\ast+\sum_{1\leq i\leq k,i\neq j}w_i(X_i)}\right]
%	\geq\prod_{i=1}^k \left(1-\frac{1}{w_i^\ast}\right),
%	\end{equation}
%	implying that Algorithm~\ref{alg:MTM-IS with different proposal distribution}  converges slower than the corresponding sequential IMH sampler.

%The condition in Theorem~\ref{final ineq} implies that .
%This condition is mainly to apply the pre-constructed proof techniques. It 
%This condition can be slightly weakened to the existence of a sequence  $\{x_n, \ n\in\mathbb{N}\}$ such that $w_j(x_n)$ converges to $w^\ast_j$, $\forall j=1,\ldots,k$.

\begin{example}\label{exam:multi_MTM}
We conducted a few simulations to examine convergence behaviors  of Algorithm~\ref{alg:MTM-IS with different proposal distribution} and the corresponding {sequential IMH} sampler at the same computational cost. 
As shown in Figure~\ref{fig:simulation for MTM with differing proposal}, we considered target densities of the form of a mixture of two standard distributions with various dimensions.
Top  plots in Figure~\ref{fig:simulation for MTM with differing proposal} correspond to Gaussian mixture targets,  $\pi=\frac{1}{2}\mathcal{N}(0,\mathbb{I}_d)+\frac{1}{2}\mathcal{N}(\vec{3},\mathbb{I}_d)$, with $d$=3, 4, and 5, respectively. Two different proposal distributions are employed: $p_1=\mathcal{N}(0,\mathbb{I}_d)$ and $p_2=\mathcal{N}(0,9\mathbb{I}_d)$. During one iteration of the MTM-IS($k$) algorithm, $k/2$ trials are independently drawn from of $p_1$, and another $k/2$ trials from $p_2$.
%with half from $p_1$ and the other half from $p_2$.
%For the bottom  plots,
%in Figure~\ref{fig:simulation for MTM with differing proposal}, 
%we investigate more heavy-tailed distributions. We
The bottom plots correspond to $t$-mixture distributions,  $\pi=\frac{1}{2}t_3(0)+\frac{1}{2}t_3(\vec{4})$, for $d$=1, 2, and 3. Two different proposal distributions are: $p_1=t_3(0)$ and $p_2=t_5(0)$, and the same implementation of MTM-IS($k$) as the previous case is employed. These plots show that Algorithm~\ref{alg:MTM-IS with different proposal distribution} and its corresponding \emph{sequential IMH} sampler differ very little in their convergence rates although theoretically we cannot claim one is necessarily better than the other without 
%When 
condition \eqref{eq:condition on MTM with differing proposal}.
All simulations are based on $10^6$ iterations on an Apple M2 chip with 16GB memory, each taking a few minutes. 
%During one iteration, all trials are independent with half from $p_1$ and the other half from $p_2$.
%Overall simulation studies suggest that when Theorem~\ref{final ineq} fails, the difference between Algorithm~\ref{alg:MTM-IS with different proposal distribution} and the sequential IMH is very subtle. It is still not definitive which one converges faster.

\begin{figure}[htbp]
	\centering
	\begin{minipage}[t]{0.3\linewidth}
		\centering
		\includegraphics[width=1\linewidth,height=1\linewidth]{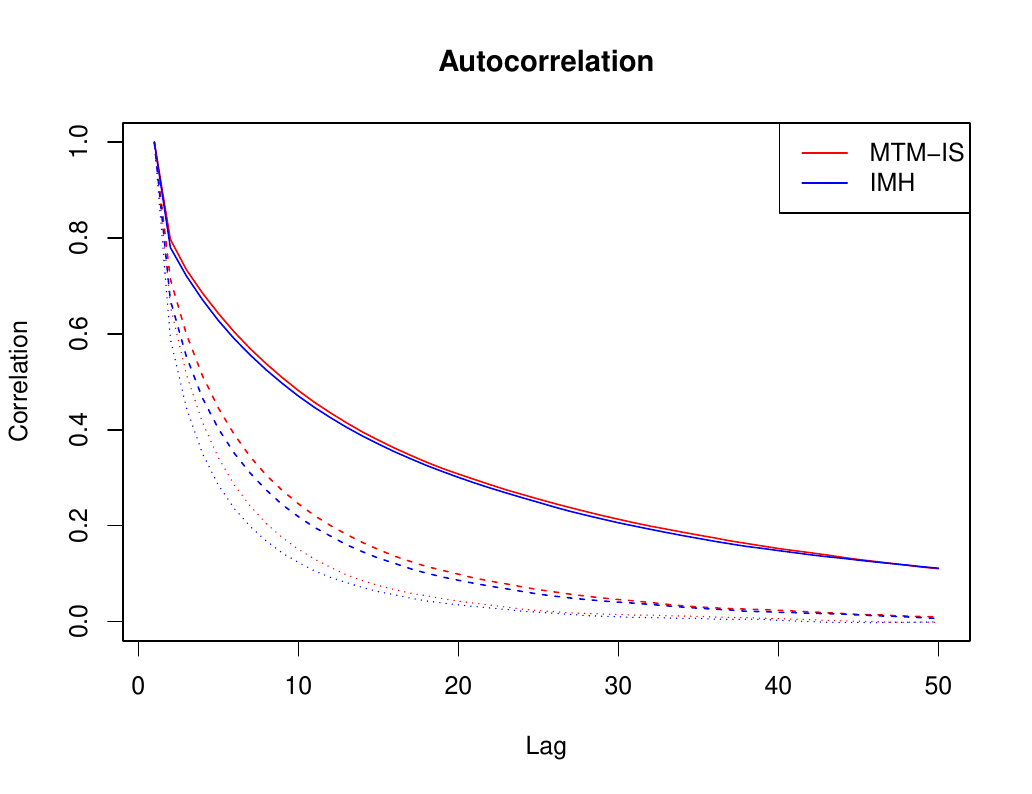}
	\end{minipage}
	\begin{minipage}[t]{0.3\linewidth}
		\centering
		\includegraphics[width=1\linewidth,height=1\linewidth]{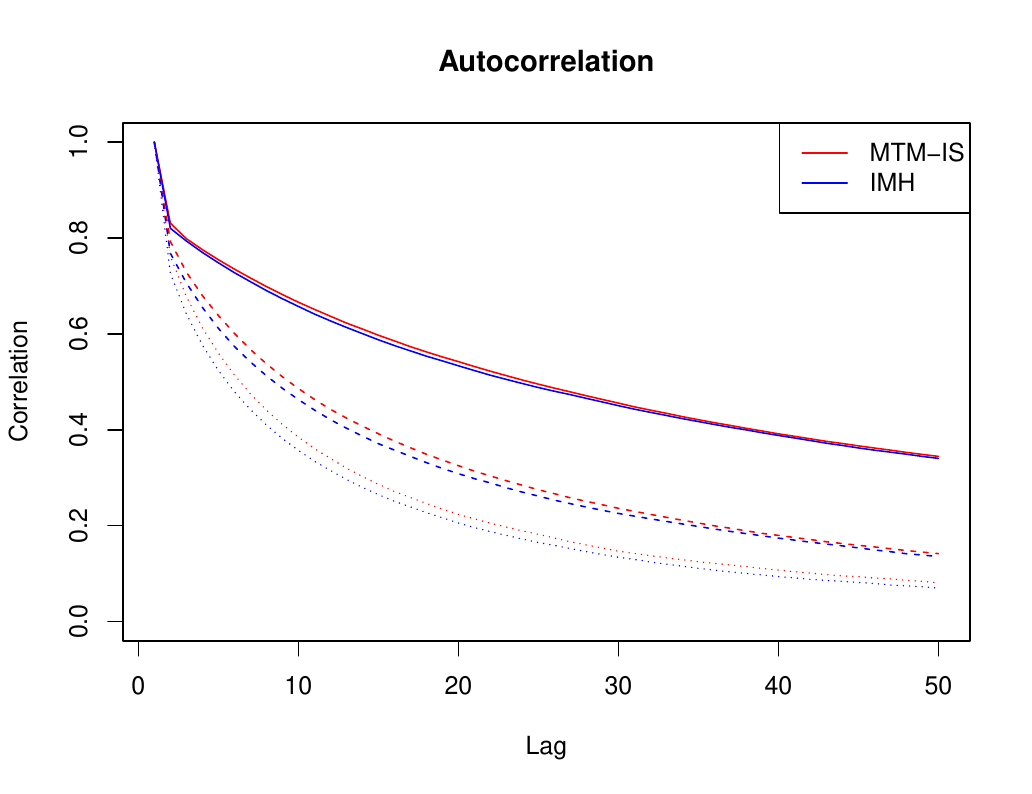}
	\end{minipage}
	\begin{minipage}[t]{0.3\linewidth}
		\centering
		\includegraphics[width=1\linewidth,height=1\linewidth]{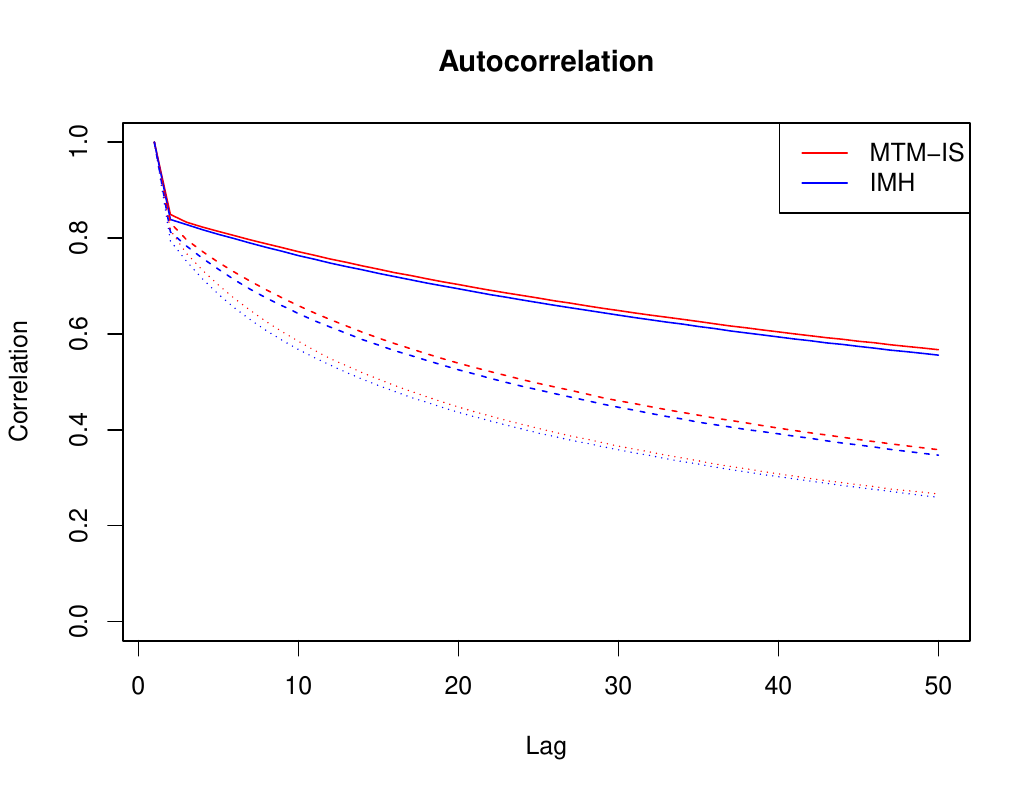}
	\end{minipage}
	
	\begin{minipage}[t]{0.3\linewidth}
		\centering
		\includegraphics[width=1\linewidth,height=1\linewidth]{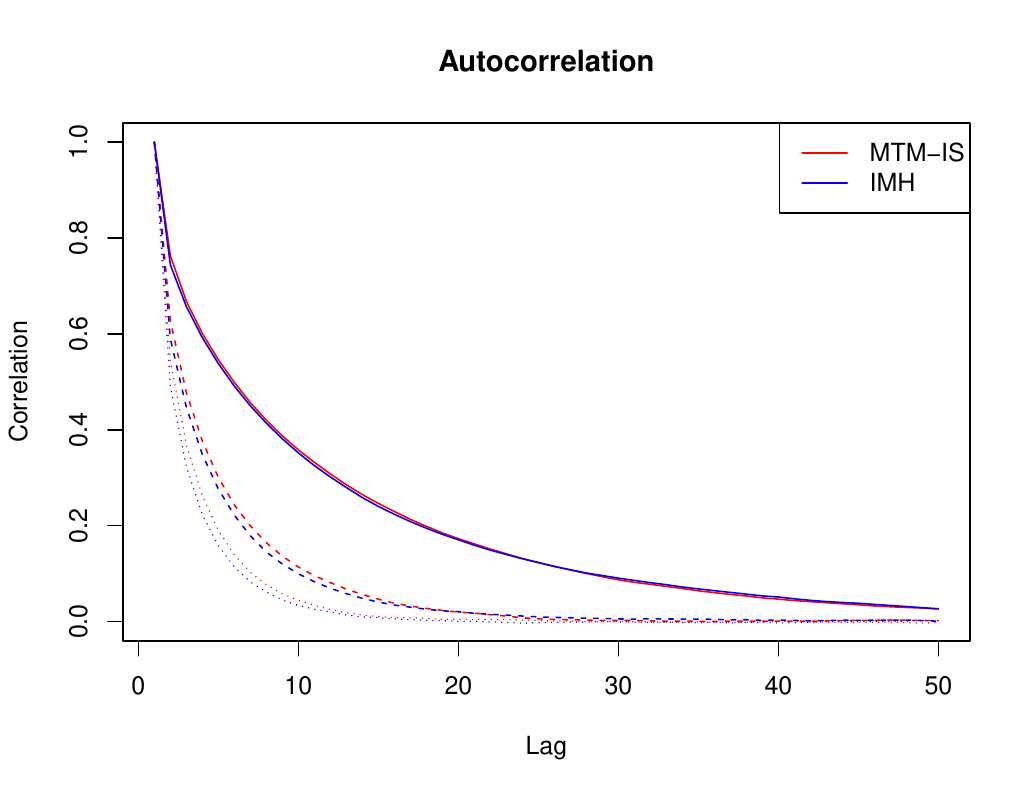}
	\end{minipage}
	\begin{minipage}[t]{0.3\linewidth}
		\centering
		\includegraphics[width=1\linewidth,height=1\linewidth]{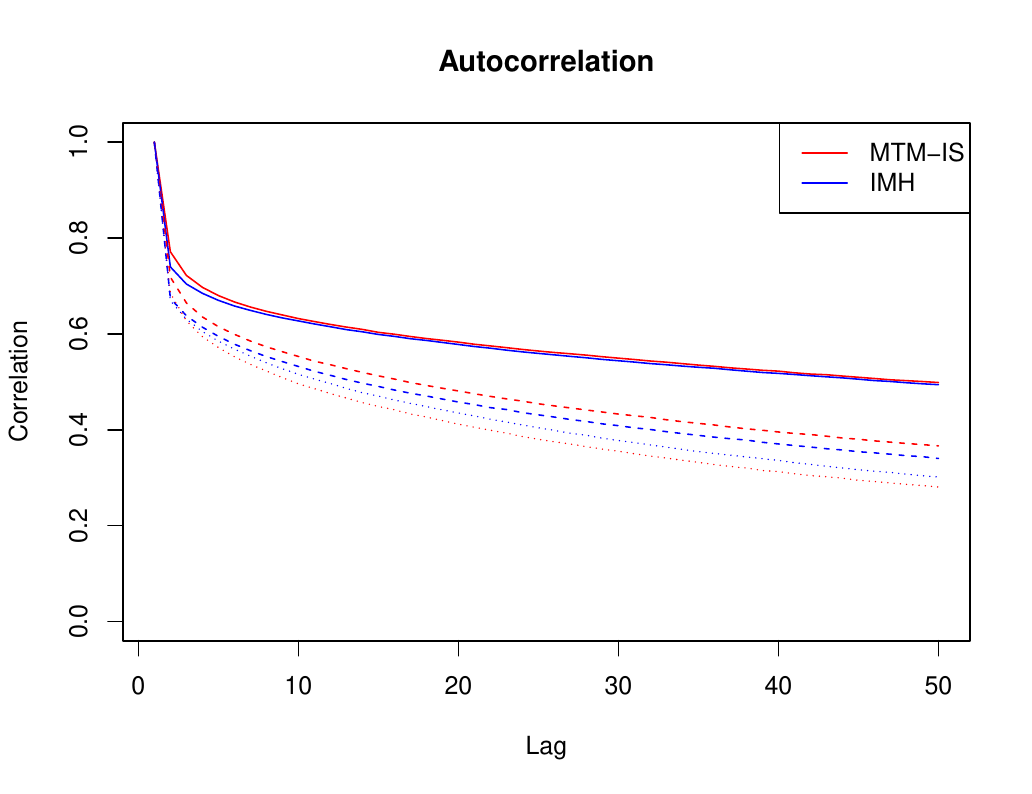}
	\end{minipage}
	\begin{minipage}[t]{0.3\linewidth}
		\centering
		\includegraphics[width=1\linewidth,height=1\linewidth]{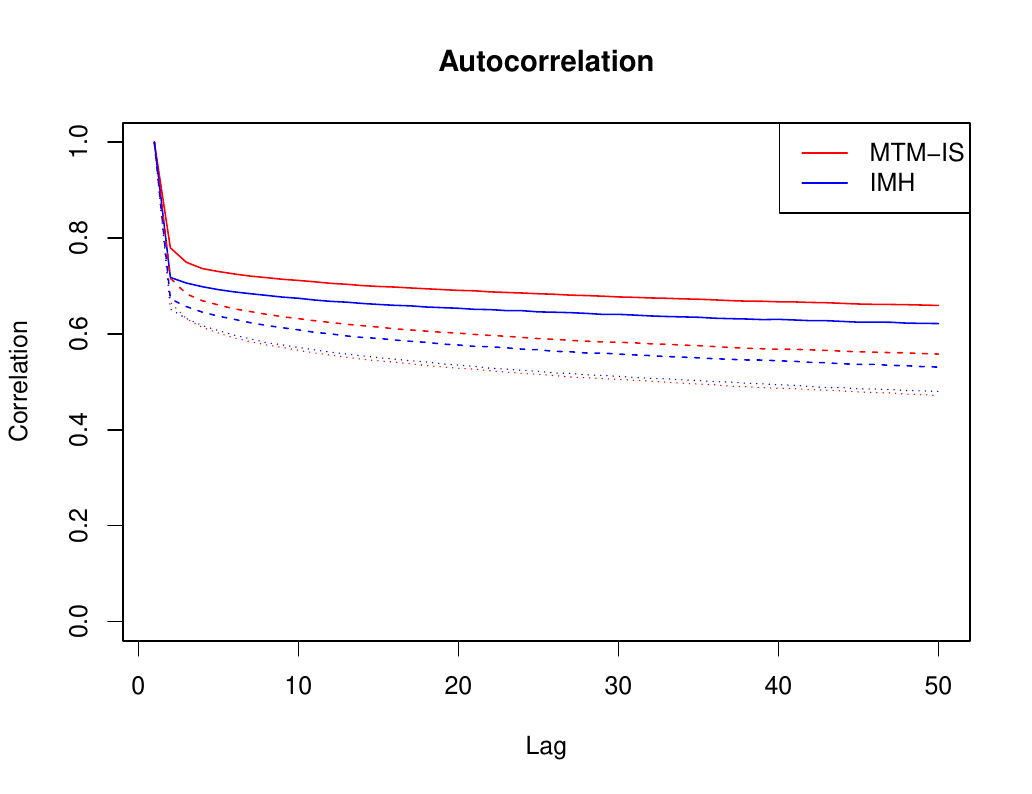}
	\end{minipage}
	\caption{\textbf{Top:} Auto-correlation plots for the Gaussian mixture targets  in Example~\ref{exam:multi_MTM}
	%simulated chains with $\pi(x)$ as a $2$-component mixture Gaussian and two Gaussian proposals with different standard deviations. 
	From left to right:  dimension $d$= $3,4,5$, respectively. 
\textbf{Bottom:} Auto-correlation plot for the $t$-mixture targets in Example~\ref{exam:multi_MTM} From left to right:  dimension $d=1,2,3$, respectively.
Solid lines: $k=2$; dashed lines: $k=6$; dotted lines: $k=10$.}
	\label{fig:simulation for MTM with differing proposal}
\end{figure}

%We conclude an empirical observation here: , it is still unknown which method converges faster in theory.

\end{example}

\subsection{A general framework}\label{sec:framework}
%Despite partitioning the state space $\bx=(\bx^a,\bx^b)$, the rest of this whole section devotes to extending the framework of simultaneously proposing multiple trials.

Inspired by the  variants of MTM just discussed,
%: multiple correlated trials (Section~\ref{sec:partialMTM}, \ref{sec:correlatedMT} and \cite{craiu2007acceleration}) and independent non-identically distributed multiple trials (Section~\ref{sec:diff_prop} and \cite{casarin2013interacting}), 
we propose a general framework to combine these variants in Algorithm~\ref{Generalized Multiple-Try Metropolis}. 
With $\pi(\cdot)$ as the target distribution on $\mathcal{X}$, we let $p(x,\by)$ denote  the proposal transition function 
%(a distribution function of $\by$ conditional on $x$) 
for multiple correlated proposals, where  $x\in\mathcal{X}$ and $\by=(y_1,\ldots,y_k)\in\mathcal{X}^k$.  
We further write the $j$-th marginal of $p(x,\by)$ as $p_j(x,y_j)=\int p(x,\by)\mathrm{d}\by_{(-j)}$,  
 and define the  $j$th {generalized importance weight} as
\begin{equation}\label{generalized importance weight}
w_j(y\mid x)=\frac{\pi(y)}{p_j(x,y)}\lambda_j(x,y),
\end{equation}
for $j=1,\ldots,k$,
where $\lambda_j$ is a symmetric function. Assuming the current state is $x$, the updating rule is summarized in Algorithm \ref{Generalized Multiple-Try Metropolis}.

\begin{algorithm}
	\caption{Generalized MTM. Suppose current state is at $x$.}\label{Generalized Multiple-Try Metropolis}
	\begin{algorithmic}[1]
		\State Draw multiple trials $y_1,\ldots,y_k$ jointly from $p(x,\by)$. Then compute $w_j(y_j\mid x)$ by \eqref{generalized importance weight} for $j=1,\ldots,k$.
		\State Select index $J$ with probability proportional to $w_j(y_j,x)$ and define $y=y_J$.
		\State Draw $x_1^\ast,x_2^\ast,\ldots,x_{J-1}^\ast,x_{J+1}^\ast,\ldots,x_k^\ast$ from the conditional distribution of $p(y,\bx^\ast)$ conditioned on $J$-th variable as $x$. And set $x_{J}^\ast=x$.
		\State Accept $y$ with the ratio $\rho=\min\left\{1,\dfrac{w_J(y\mid x)+\sum_{i\neq J}w_i(y_i\mid x)}{w_J(x\mid y)+\sum_{i\neq J}w_i(x_i^\ast\mid y)}\right\}$.
	\end{algorithmic}
\end{algorithm}

%There are also semi-deterministic variants of \textit{Multiple-Try Metropolis} (e.g., the griddy Gibbs \textit{Multiple-Try Metropolis} in \cite{liu2000multiple} and Hamiltonian Monte Carlo \textit{Multiple-Try Metropolis} in \cite{qin2001multipoint}). But they are more difficult to analyse due to their deterministic aspects and are omitted here.

If we require $p(x,\by)$ to have the same marginals for different $y_j$'s, the algorithm reduces to that of \cite{craiu2007acceleration}; if we require $p(x;\by)$ to be independent among the $y_j$'s, it reduces to that of \cite{casarin2013interacting}. Note that the balancing proposals are drawn to facilitate the computation of $\rho$, and this guarantees the detailed balance of the  MTM design. The following result is expected and its detailed proof is deferred to the Appendix.

\begin{theorem}\label{detailed balance}
	The generalized MTM transition rule (Algorithm \ref{Generalized Multiple-Try Metropolis}) satisfies the detailed balance condition and hence induces a reversible Markov chain with $\pi$ as its invariant distribution.
\end{theorem}

Defining $\bx^\ast(j)\triangleq(x^\ast_1,\ldots,x^\ast_{j-1},x,x^\ast_{j+1},x^\ast_k)$, one can determine the transition density of the generalized MTM framework via the same spirit employed in the proof of Theorem~\ref{thm:H}:
\begin{align*}
A(x,y)=&\pi(y) \sum_{j=1}^k 
\Bigg[ p_j(x,y)p_j(y,x) \lambda_j(x,y) \times  \\
&   \int u_j(\bx^\ast(j),\by) p(x,\by_{(-j)} \mid y_j=y) 
p(y,\bx^\ast_{(-j)}\mid x_j^\ast=x)\prod_{i\neq j}\mathrm{d}y_i\mathrm{d}x_i^\ast \Bigg],
\end{align*} 
where we write $u_j(\bx,\by)\triangleq\min\left\{\left(\sum_{i=1}^k w_i(y_i\mid x_j)\right)^{-1},\left(\sum_{i=1}^kw_i(x_i\mid y_j)\right)^{-1}\right\}$ 
for any $\bx=(x_1,\ldots,x_k)$ and $\by=(y_1,\ldots,y_k)$. A detailed derivation of this formula can be found in the proof of Theorem \ref{detailed balance}. 
%For here, we only make illustrations based on it. Roughly speaking, unless more conditions are imposed, we cannot provide  tight bounds for this transition function. This is not too surprising as we are so far still unable to provide sharp bounds for the  {Random-walk Metropolis-Hastings} algorithm (see conjectures in \cite{atchade2007geometric} and  discussions in \cite{wang2020exact}).

As demonstrated in Algorithms~\ref{alg:MTM-SRSWOR}, \ref{alg:MTM-SRSWOR2} and \ref{alg:MTM-IS with different proposal distribution}, we find that sometimes we do not need to draw balancing trials for MTM to retain the detailed balance.
A natural question then arises: can we find a general condition under which which  MTM can avoid the drawing of balancing trials? The following theorem provides a sufficient condition that covers all the cases we discussed.

\begin{theorem}\label{thm:condition}
	If, for any pair $(x,y)$ and $\forall j$, the joint proposal distribution satisfies
	\begin{equation}\label{condition}
	p(x,\by_{(-j)}\mid y_j=y)=p(y,\by_{(-j)}\mid y_j=x),
	\end{equation}
we can  maintain the detailed balance by setting $x^\ast_j\triangleq y_j$ for $j\neq J$ in Algorithm \ref{Generalized Multiple-Try Metropolis}.
\end{theorem}
%Theorem~\ref{thm:condition} justifies why we do not need to 
%that we do not need to  draw balancing trials in 

\begin{remark}[Correlated multiple trials]
As demonstrated in Sections~\ref{sec:partialMTM} and \ref{sec:correlatedMT}, letting the proposed multiple trials be correlated (especially negatively) can be  helpful in improving the chain's convergence. A useful strategy is to use multiple trials as stepping stones to move from one mode of the distribution to another, similar in spirit to Hamiltonian/hybrid Monte Carlo \citep{qin2001multipoint,liu2008monte} and the griddy Gibbs MTM \citep{liu2000multiple}. Indeed, it was shown empirically in \cite{qin2001multipoint} that applying MTM to HMC trajectories may further improve the sampling efficiency.
However, an in-depth theoretical analysis as carried out here is much more challenging due to the semi-deterministic nature of aforementioned algorithms.
\end{remark}

\begin{remark}[Employing multiple distributions in MTM]
Intuitively, one may hope that using different distributions for each trial could help us explore the  state space better. Our results in Section~\ref{sec:diff_prop}, however, demonstrate that it is still not very useful  under the IMH framework if the multiple trials are independent. It may be helpful for the partial MTM framework discussed in Section~\ref{sec:partialMTM}.
\end{remark}

\section{Concluding Remarks}\label{sec 5}
We have presented a complete eigen-decomposition and  convergence rate analysis for  the MTM-IS, and compared it with the ``thinned" IMH sampler (of the same computational cost). With the exact form of  eigenvalues of the MTM-IS, we proved rigorously that the 
%Multiple-Try Metropolis Independent 
sampler is not as efficient as the simpler ``thinned" IMH approach. To the best of our knowledge, this is the first exact rate result known  for a MTM type algorithm, although the result's implication is less than encouraging. A good news is that, in a more realistic setting of MTM applications as explained in Section~\ref{sec:partialMTM}, we can show that MTM improves upon the standard IMH and does not have a suitable competitor. 

In a quest for finding  advantages MTM may offer, we consider a slightly modified framework that encompasses a few variants of MTM published in the literature. We found that even under the  IMH framework, it is possible to construct a MTM algorithm, using either  stratified sampling or partial sampling, or sampling without replacement, to gain efficiency. A key to such efficiency gain is to allow  multiple trials to be either more dispersed than independent ones (Section~\ref{sec 4}) or applied only to certain ``low-cost'' parts (Section~\ref{sec:partialMTM}). Detailed theoretical understanding and guiding principles, however, are still lacking and awaiting further endeavors.

\backmatter

\bmhead{Acknowledgments}
We thank the National Science Foundation of the United States (DMS-1903139 and DMS-2015411) for partially supporting the research. Part or of work was done when Yang was a student in the
{School of Gifted Young}, \orgname{University of Science and Technology of China}.

\section*{Declarations}

The authors have no competing interests that are directly or indirectly related to the work submitted for publication.

\iffalse
Some journals require declarations to be submitted in a standardised format. Please check the Instructions for Authors of the journal to which you are submitting to see if you need to complete this section. If yes, your manuscript must contain the following sections under the heading ‘Declarations’

\begin{itemize}
\item Funding
\item Conflict of interest/Competing interests (check journal-specific guidelines for which heading to use)
\item Ethics approval 
\item Consent to participate
\item Consent for publication
\item Availability of data and materials
\item Code availability 
\item Authors' contributions
\end{itemize}

\noindent
If any of the sections are not relevant to your manuscript, please include the heading and write `Not applicable' for that section. 

\bigskip
\begin{flushleft}%
Editorial Policies for:

\bigskip\noindent
Springer journals and proceedings: \url{https://www.springer.com/gp/editorial-policies}

\bigskip\noindent
Nature Portfolio journals: \url{https://www.nature.com/nature-research/editorial-policies}

\bigskip\noindent
\textit{Scientific Reports}: \url{https://www.nature.com/srep/journal-policies/editorial-policies}

\bigskip\noindent
BMC journals: \url{https://www.biomedcentral.com/getpublished/editorial-policies}
\end{flushleft}
\fi

\begin{appendices}

\section{Detailed Proofs}\label{secA1}

\begin{proof}[Proof of Theorem~\ref{spectrum thm}]
Before proving the theorem, we first define the following   additional notations and concepts.
Let $A(\cdot,\cdot)$ denote the Markov transition kernel implied  by our algorithm. The operator $K$  associated with the resulting Markov chain is defined as follows:  for any measurable function $f$ defined on $\mathcal{X}$, operator $K$ maps $f$ to another function defined on $\cal X$:
\begin{equation*}
Kf(x)=\int_{\mathcal{X}}f(y)A(x,\mathrm{d}y).
\end{equation*}
We require that  function $f\in L^2(\pi)$. It is easy to see that $Kf\in L^2(\pi)$ as well, meaning that $K$ defines a linear bounded operator on the Hilbert space $L^2(\pi)$ with operator norm $1$. 
For any set $S\subset\mathcal{X}$, we shall also denote $\chi_S:\mathcal{X}\rightarrow\{0,1\}$ as the indicator function which equals $1$ if and only if on $S$. 
Intuitively, $K$ is just a conditional expectation operator. Note that the constant function $1$ is automatically an eigenfunction of eigenvalue $1$. We are interested in finding  the spectral gap, i.e., the difference between 1 and the second largest eigenvalue. 
We thus focus on the restricted operator $K_0$ defined on the orthogonal complement of the constant function:
\begin{equation*}
L^2_0(\pi)=\left\{f\in L^2(\pi):\int_{\mathcal{X}}f(x)\mathrm{d}x=0\right\}.
\end{equation*}

Given Theorem~\ref{thm:H}, we divide the operator $K_0$ into two parts: $\forall f\in L^2_0(\pi)$, 
\begin{align*}
K_0 f(x)&=R(x)f(x)+\int_{\mathcal{X}}\min\left[H[w(x)],H[w(y)]\right]f(y)\pi(y)\mathrm{d}y=:M_Rf(x)+Uf(x).
\end{align*}
Before presenting the formal proof, we remark that this decomposition has the same nature as that in Section 2.1 of \cite{liu1996metropolized}, in which the multiplication operator $M_R$ is a low-rank component and the integral-like operator $U$ that resembles  the upper triangular matrix in the discrete case. This proof is analogous to that in \cite{atchade2007geometric}. The formal proof is divided into the following steps.

\vspace{2mm}
	
	\noindent{\bf Step 1.}
	    We first show that operator $U$ is compact. Under the following condition,
		\begin{equation*}
		\int_{\mathcal{X}}\int_{\mathcal{X}}\min\left\{H[w(x)],H[w(y)]\right\}^2\pi(x)\pi(y)\mathrm{d}x\mathrm{d}y<\infty,
		\end{equation*}
		operator $U$ is Hilbert-Schmidt, and therefore compact. Hence, by Weyl's perturbation theorem, we have 
		\begin{equation*}
		\sigma_{\text{ess}}(K_0)=\sigma_{\text{ess}}(M_R)\subset \text{ess-ran}(R).
		\end{equation*}
		
	\noindent{\bf Step 2.}
	    Given this, combined with the decomposition
		\begin{equation*}
		\sigma(K_0)=\sigma_{\text{ess}}(K_0)\cup\sigma_d(K_0),
		\end{equation*}
		we know that it suffices to prove that $\sigma_d(K_0)\subset \text{ess-ran}(R)$, i.e. all eigenvalues of $K_0$ are in the essential range of $R$. To proceed, we assume that there exists $f_0\in L^2_0(\pi)$ and $\lambda\notin \text{ess-ran}(R)$, but $K_0f_0=\lambda f_0$.
		
		Direct computations yield that for any $f\in L^2_0(\pi)$
		\begin{align*}
		Uf(x)&=\int\min\left\{H[w(x)],H[w(y)]\right\}f(y)\pi(y)\mathrm{d}y\\
		&=\int_{\{y:w(y)\geq w(x)\}}H[w(y)]f(y)\pi(y)\mathrm{d}y+\int_{\{y:w(y)< w(x)\}}H[w(x)]f(y)\pi(y)\mathrm{d}y\\
		&=\int_{\{y:w(y)\geq w(x)\}}\left\{H[w(y)]-H[w(x)]\right\}f(y)\pi(y)\mathrm{d}y.
		\end{align*}
		Since we assume that $\lambda\notin \text{ess-ran}(R)$, we have $\kappa=\text{ess}\inf\left(\mid  R(x)-\lambda\mid\right)>0$. We can rearrange  equation $K_0f_0=\lambda f_0$ to arrive at
		\begin{equation}
		\int_{\{y:w(y)\geq w(x)\}}\frac{H[w(x)]-H[w(y)]}{R(x)-\lambda}f_0(y)\pi(y)\mathrm{d}y=f_0(x),
		\end{equation}
		which can be simplified as $Nf_0=f_0$ with $N$ being an operator  well-defined on $L^2(\pi)$ (rather than $L^2_0(\pi)$ in which $K_0$ is defined). Then, we aim to derive a contradiction about the spectral radius $\text{radii}(G)\triangleq\sup\{\mid\lambda\mid:\lambda\in\sigma(G)\}$ for some linear operator $G$ on $L^2(\pi)$ induced by $N$.
		
		\vspace{2mm}
		
		\noindent{\bf Step 3.}
		Since $f_0$ is not identically vanishing,  we can find $u<w^\ast$ so that $f_0$ is not null on $\{x\in\mathcal{X}:u<w(x)\leq w^\ast\}$. For any partition $I_n=(u=u_n\leq u_{n-1}\leq\ldots\leq u_0=w^\ast)$, we denote $D_i=\{x\in\mathcal{X}:u_i<w(x)\leq u_{i-1}\}$ and $L^2_i(\pi)=\{h\in L^2_0(\pi):h(x)=0,\forall x\notin D_i\}$ for $i=1,\ldots,n$. Then $L^2_i(\pi)$ is a closed subspace of $L^2_0(\pi)$, thus a Hilbert space. Moreover, we introduce $M_{D_i}$ as the restriction operator onto $D_i$ on $L^2(\pi)$, by letting $M_{D_i}g(x)=\chi_{D_i}(x)g(x)$ for any $g\in L^2(\pi)$.
		
		We know that
		\begin{align*}
		M_{D_1}Nf_0(x)&=\int_{\{y:w(y)\geq w(x)\}}\frac{H[w(x)]-H[w(y)]}{R(x)-\lambda}f_0(y)\chi_{D_1}(x)\pi(y)\mathrm{d}y\\
		&=\int_{\{y:w(y)\geq w(x)\}}\frac{H[w(x)]-H[w(y)]}{R(x)-\lambda}\chi_{D_1}(y)f_0(y)\chi_{D_1}(x)\pi(y)\mathrm{d}y\\
		&=M_{D_1}NM_{D_1}f_{0},
		\end{align*}
		where the second inequality follows from the fact that $y\notin D_1$ and $w(y)\geq w(x)$ would together imply that $x\notin D_1$. Obtaining from $Nf_0=f_0$ and $M^2_{D_1}=M_{D_1}$, we then have $f_{0,D_1}\triangleq M_{D_1}f_0=M_{D_1}Nf_0=M_{D_1}NM_{D_1}f_{0}=M_{D_1}NM_{D_1}f_{0,D_1}$.
		
		In the same manner, we have
		\begin{align*}
		&M_{D_i}Nf_0(x)\\
		=&\int_{\{y:w(y)\geq w(x)\}}\frac{H[w(x)]-H[w(y)]}{R(x)-\lambda}f_0(y)\chi_{D_i}(x)\pi(y)\mathrm{d}y\\
		=&\sum_{k=1}^{i-1}\int_{\{y\in D_k\}}\frac{H[w(x)]-H[w(y)]}{R(x)-\lambda}f_0(y)\chi_{D_i}(x)\pi(y)\mathrm{d}y\\
		&+\int_{\{y:u_{i-1}>w(y)\geq w(x)\}}\frac{H[w(x)]-H[w(y)]}{R(x)-\lambda}f_0(y)\chi_{D_i}(x)\pi(y)\mathrm{d}y\\
		=&M_{D_i}h_i(x) + \int_{\{y:u_{i-1}>w(y)\geq w(x)\}}\frac{H[w(x)]-H[w(y)]}{R(x)-\lambda}\chi_{D_i}(y)f_0(y)\chi_{D_i}(x)\pi(y)\mathrm{d}y\\
		=&M_{D_i}h_i(x)+M_{D_i}NM_{D_i}f_{0,D_i},
		\end{align*}
		where $f_{0,D_i}\triangleq M_{D_i}f_0$ and
		\begin{equation}\label{eq:define h_i}
		h_i(x)=\sum_{k=1}^{i-1}\int_{\{y\in D_k\}}\frac{H[w(x)]-H[w(y)]}{R(x)-\lambda}f_0(y)\pi(y)\mathrm{d}y.
		\end{equation}
		Rearranging these formulae, we know that
		\begin{align}\label{eqset}
		M_{D_1}NM_{D_1}f_{0,D_1}&=f_{0,D_1},\\
		M_{D_2}NM_{D_2}f_{0,D_2}&=f_{0,D_2}-M_{D_2}h_2,\\
		&\ldots\notag\\
		M_{D_n}NM_{D_n}f_{0,D_n}&=f_{0,D_n}-M_{D_n}h_n.
		\end{align}

		We claim that \eqref{eqset} implies  that $\text{radii}(M_{D_i}NM_{D_i})\geq1$ holds true for at least one index $i\in\{1,\ldots,n\}$.
		Assuming the converse is true, then $M_{D_1}NM_{D_1}f_{0,D_1}=f_{0,D_1}$ implies that $f_{0,D_1}=0$ (since 1 cannot be an eigenvalue of $M_{D_1}NM_{D_1}$). Consequently, $h_2=0$ follows automatically from its definition \eqref{eq:define h_i}, and $M_{D_2}NM_{D_2}f_{0,D_2}=f_{0,D_2}$ implies that $f_{0,D_2}=0$. This argument can be carried out recursively until $n$, indicating that $f_0$ has to vanish on $\{x\in\mathcal{X}:u<w(x)\leq\bar{w}\}$, resulting in a contradiction!
		
		\vspace{2mm}
		
		\noindent{\bf Step 4.}
		Finally, we  show that for sufficiently small increments, we can make
		\begin{equation*}
		\text{radii}(M_{D_i}NM_{D_i})<1, \ \forall i.
		\end{equation*}
		First, the mapping\begin{equation*}
		H:v\in\mathbb{R}_+\mapsto k\underbrace{\int_{\mathcal{X}}\ldots\int_\mathcal{X}}_{k-1}\frac{1}{v+\sum_{j=1}^{k-1} w(y_j)}\prod_{j=1}^{k-1} p(x)\mathrm{d}x
		\end{equation*}
		is continuous, at least on $[u,\bar{w}]$.
		
		Second, $\forall g\in L^2_i(\pi)$ with $\Vert g\Vert=1$, by the Cauchy-Schwarz inequality we have
		\begin{align*}
		&\Vert M_{D_i}NM_{D_i}g\Vert^2\\
		=&\int_{\{y\in D_i\}} \left\{ \int_{\{y:w(y)\geq w(x)\}}\frac{H[w(x)]-H[w(y)]}{R(x)-\lambda}g(y)\chi_{D_i}(x)\pi(y)\mathrm{d}y\right\}^2\pi(x)\mathrm{d}x\\
		\leq&\left(\frac{\max_{[u_i,u_{i-1}]}H-\min_{[u_i,u_{i-1}]}H}{\kappa}\right)^2\int_{\{y\in D_i\}}g^2(y)\pi(y)\mathrm{d}y\\
		\leq&\left(\frac{\max_{[u_i,u_{i-1}]}H-\min_{[u_i,u_{i-1}]}H}{\kappa}\right)^2=\left(\frac{\text{osc}_{[u_i,u_{i-1}]}H}{\kappa}\right)^2,
		\end{align*}
		where $\text{osc}_{[u_i,u_{i-1}]}H\triangleq\max_{[u_i,u_{i-1}]}H-\min_{[u_i,u_{i-1}]}H$ denotes the oscillation of $H$ within $[u_i,u_{i-1}]$.
		Therefore, $\Vert M_{D_i}NM_{D_i}\Vert\leq\text{osc}_{[u_i,u_{i-1}]}H/\kappa$. At last, if we choose the partition to be sufficiently small, we would have $\text{radii}(M_{D_i}NM_{D_i})<1$ for all $i$. We then derive a final contradiction to assert that $\sigma_d(K_0)\subset \text{ess-ran}(R)$, ending the proof.
\end{proof}

\begin{proof}[Proof of Theorem~\ref{comparison thm}]
	In this proof, every random variable $X$ is taken independently from $p$. This inequality is proved by induction. First, for $k=1$, the inequality reduces to equality due to a previous result of \cite{liu1996metropolized} and \cite{atchade2007geometric}. For $k=2$, we see that
	\begin{align*}
	&1-\mathbb{E}\left[\frac{2}{w^\ast+w(X)}\right]-\left(1-\frac{1}{w^\ast}\right)^2\\
	=&\frac{1}{w^\ast}\mathbb{E}\left[\frac{2w(X)}{w^\ast+w(X)}-\frac{1}{w^\ast}\right]
	\geq \frac{1}{w^\ast}\mathbb{E}\left[\frac{2w(X)}{2w^\ast}-\frac{1}{w^\ast}\right]=0.
	\end{align*} 
	For $k\geq3$, we will prove the following recursive inequality, which leads to the conclusion of the theorem:
	\begin{equation}\label{eqn:recursive}
	1-\mathbb{E}\left[\frac{k}{w^\ast+\sum_{i=1}^{k-1}w(X_i)}\right]\geq\left(1-\frac{1}{w^\ast}\right)\left(1-\mathbb{E}\left[\frac{k-1}{w^\ast+\sum_{i=1}^{k-2}w(X_i)}\right]\right).
	\end{equation}
	We prove by simply computing the difference between the two sides:
	\begin{align*}
	&1-\mathbb{E}\left[\frac{k}{w^\ast+\sum_{i=1}^{k-1}w(X_i)}\right]-\left(1-\frac{1}{w^\ast}\right)\left(1-\mathbb{E}\left[\frac{k-1}{w^\ast+\sum_{i=1}^{k-2}w(X_i)}\right]\right)\\
	=&\mathbb{E}\left[\frac{k-1}{w^\ast+\sum_{i=1}^{k-2}w(X_i)}\right]+\frac{1}{w^\ast}-\mathbb{E}\left[\frac{k}{w^\ast+\sum_{i=1}^{k-1}w(X_i)}\right]-\mathbb{E}\left[\frac{k-1}{w^\ast[w^\ast+\sum_{i=1}^{k-2}w(X_i)]}\right]\\
	=&\underbrace{(k-1)\left(\mathbb{E}\left[\frac{1}{w^\ast+\sum_{i=1}^{k-2}w(X_i)}\right]-\mathbb{E}\left[\frac{1}{w^\ast+\sum_{i=1}^{k-1}w(X_i)}\right]\right)}_{(i)}\\
	&+\underbrace{\frac{1}{w^\ast}-\mathbb{E}\left[\frac{1}{w^\ast+\sum_{i=1}^{k-1}w(X_i)}\right]-\mathbb{E}\left[\frac{k-1}{w^\ast[w^\ast+\sum_{i=1}^{k-2}w(X_i)]}\right]}_{(ii)}.
	\end{align*}
	We note that $(i)$ can be modified as
	\begin{align*}
	(i)=&(k-1)\left(\mathbb{E}\left[\frac{1}{w^\ast+\sum_{i=1}^{k-2}w(X_i)}\right]-\mathbb{E}\left[\frac{1}{w^\ast+\sum_{i=1}^{k-1}w(X_i)}\right]\right)\\
	=&\sum_{j=1}^{k-1}\left(\mathbb{E}\left[\frac{1}{w^\ast+\sum_{1\leq i\leq k-1,i\neq j}w(X_i)}\right]-\mathbb{E}\left[\frac{1}{w^\ast+\sum_{i=1}^{k-1}w(X_i)}\right]\right)\\
	=&\sum_{j=1}^{k-1}\mathbb{E}\left\{\frac{w(X_j)}{[w^\ast+\sum_{1\leq i\leq k-1,i\neq j}w(X_i)][w^\ast+\sum_{i=1}^{k-1}w(X_i)]}\right\}.
	\end{align*}For $(ii)$, we have
	\begin{align*}
	(ii)=&\mathbb{E}\left\{\frac{\sum_{j=1}^{k-1}w(X_j)}{w^\ast[w^\ast+\sum_{i=1}^{k-1}w(X_i)]}\right\}-\mathbb{E}\left[\frac{k-1}{w^\ast[w^\ast+\sum_{i=1}^{k-2}w(X_i)]}\right]\\
	=&\sum_{j=1}^{k-1}\mathbb{E}\left\{\frac{w(X_j)}{w^\ast[w^\ast+\sum_{i=1}^{k-1}w(X_i)]}\right\}-\sum_{j=1}^{k-1}\mathbb{E}\left\{\frac{w(X_j)}{w^\ast[w^\ast+\sum_{1\leq i\leq k-1,i\neq j}w(X_i)]}\right\}\\
	=&-\sum_{j=1}^{k-1}\mathbb{E}\left\{\frac{w(X_j)^2}{w^\ast[w^\ast+\sum_{i=1}^{k-1}w(X_i)][w^\ast+\sum_{1\leq i\leq k-1,i\neq j}w(X_i)]}\right\}.
	\end{align*}
	In conclusion, we have
	\begin{align*}
	&1-\mathbb{E}\left[\frac{k}{w^\ast+\sum_{i=1}^{k-1}w(X_i)}\right]-\left(1-\frac{1}{w^\ast}\right)\left(1-\mathbb{E}\left[\frac{k-1}{w^\ast+\sum_{i=1}^{k-2}w(X_i)}\right]\right)=(i)+(ii)\\
	=&\sum_{j=1}^{k-1}\mathbb{E}\left\{\frac{w(X_j)[w^\ast-w(X_j)]}{w^\ast[w^\ast+\sum_{i=1}^{k-1}w(X_i)][w^\ast+\sum_{1\leq i\leq k-1,i\neq j}w(X_i)]}\right\}\geq0.
	\end{align*}
	Consequently, suppose the inequality \eqref{comparison inequality} holds for $k-1$, i.e.,
	\begin{equation*}
	    1-\mathbb{E}\left[\frac{k-1}{w^\ast+\sum_{i=1}^{k-2}w(X_i)}\right]\geq \left(1-\frac{1}{w^\ast}\right)^{k-1},
	\end{equation*}
	from \eqref{eqn:recursive} it immediately follows
	\begin{equation*}
	    1-\mathbb{E}\left[\frac{k}{w^\ast+\sum_{i=1}^{k-1}w(X_i)}\right]\geq\left(1-\frac{1}{w^\ast}\right)\left(1-\mathbb{E}\left[\frac{k-1}{w^\ast+\sum_{i=1}^{k-2}w(X_i)}\right]\right)\geq\left(1-\frac{1}{w^\ast}\right)^{k}
	\end{equation*}
	By induction, the final result \eqref{comparison inequality} holds for arbitrary $k\geq1$.
\end{proof}

\begin{proof}[Proof of Theorem~\ref{final ineq}]
Part 1 derives the convergence rate of Algorithm~\ref{alg:MTM-IS with different proposal distribution}. Part 2 derives the convergence rate of the corresponding sequential IMH sampler. Part 3 finishes by deriving the inequality~\eqref{eq:generalized_rate} via induction. 

\vspace{2mm}

\noindent{\bf Part 1.} Via straight forward computation, the transition probability of Algorithm~\ref{alg:MTM-IS with different proposal distribution} has the following formula $(x\neq y)$
\begin{align*}
    A(x,y)=&\sum_{j=1}^{k}\underbrace{\int\ldots\int}_{k-1}\frac{w_j(y)p_j(y)\prod_{i\neq j}p_i(y_i)\mathrm{d}y_i}{\max\{w_j(y)+\sum_{i\neq j}w_i(y_i),w_j(x)+\sum_{i\neq j}w_i(y_i)\}}\\
    =&\pi(y)\sum_{j=1}^{k}\underbrace{\int\ldots\int}_{k-1}\frac{\prod_{i\neq j}p_i(y_i)\mathrm{d}y_i}{\max\{w_j(y),w_j(x)\}+\sum_{i\neq j}w_i(y_i)}.
\end{align*}
Plug $\max\{w_j(y),w_j(x)\}\leq w_j^\ast$ into this formula to get
\begin{equation*}
    A(x,y)\geq\pi(y)\mathbb{E}_p\left[\sum_{j=1}^k\frac{1}{w_j^\ast+\sum_{1\leq i\leq k,i\neq j}w_i(X_i)}\right],
\end{equation*}
where $X_i$ is taken independently from $p_i(\cdot)$.
Actually this inequality is sufficient to derive a decomposition of $A(x,\cdot)$ as in \eqref{decomposition}. As shown in the proof of Theorem~\ref{rate theorem}, we upper bound the convergence rate by $1-\sum_{j=1}^{k}\mathbb{E}\left[\frac{1}{w_j^\ast+\sum_{i=1,i\neq j}^{k}w_i(X_i)}\right]$ via coupling argument, Lemma~\ref{lem:coupling}.

Specifically, when there exists $x^\ast$ such that $w_j(x^\ast)=w_j^\ast$ for all $j=1,\ldots,k$, we find for any $y\neq x^\ast$,
\begin{equation*}
   A(x^\ast,y)=\pi(y)\mathbb{E}_p\left[\sum_{j=1}^k\frac{1}{w_j^\ast+\sum_{1\leq i\leq k,i\neq j}w_i(X_i)}\right].
\end{equation*}
Consequently, the rejection probability at $x^\ast$ is
\begin{equation*}
    R(x^\ast)=1-\mathbb{E}_p\left[\sum_{j=1}^k\frac{1}{w_j^\ast+\sum_{1\leq i\leq k,i\neq j}w_i(X_i)}\right].
\end{equation*}
Then we lower bound the convergence rate via Lemma~\ref{lem:lower}.

\vspace{2mm}

\noindent{\bf Part 2.} Turn to the corresponding sequential IMH sampler. For simplicity, we utilize the concept of $L^2$ operators introduced in Section~\ref{sec 2} to derive upper bounds. Within one iteration, the sampler runs an interior loop of length $k$, with each step as a vanilla IMH step using proposal $p_i$. The transition probability of a vanilla IMH step is
\begin{align*}
    A^{(i)}(x,y)=\frac{1}{\max\{w_i(x),w_i(y)\}}\pi(y)+\left(1-\int_{\mathcal{X}}\frac{1}{\max\{w_i(x),w_i(y)\}}\pi(y)\mathrm{d}y\right)\delta_x(y).
\end{align*}
Denote $K^{(i)}$ as the operator defined in $L^2(\pi)$ by $K^{(i)}f(x)=\int f(y)A^{(i)}(x,y)\mathrm{d}y$, and denote $K_0^{(i)}$ as the restriction of $K^{(i)}$ onto $L^2_0(\pi)$, the orthogonal complement of the constant function of $L^2(\pi)$. Theorem~\ref{spectrum thm} implies $\Vert K_0^{(i)}\Vert\leq1-1/w_i^\ast$. Denote the whole transition probability of one iteration as $\bar{A}$ and associated operators as $\bar{K}$ and $\bar{K}_0$. Consequently,
\begin{equation*}
    \Vert \bar{K}_0\Vert_2=\Vert\bar{K}_0^{(k)}\cdots\bar{K}_0^{(1)}\Vert_2\leq\prod_{i=1}^{k}(1-1/w_i^\ast).
\end{equation*}
Let $p_n(x)= \bar{A}_n(p_0, x)$ denote the  distribution of the $n$-th state of the Markov chain after $n$ steps from initialization $p_0$. \citet{liu1995covariance} establishes
\begin{align*}
    \Vert p_n-\pi\Vert_{TV}\leq 2d_\chi (\pi,p_n) \leq 2\Vert \bar{K}_0^n \Vert_2 d_\chi (\pi,p_0).
\end{align*}
Furthermore, we obtain an upper bound on the convergence rate defined in \eqref{eq:vd_r}: $r\leq\Vert\bar{K}_0\Vert_0=\prod_{i=1}^{k}(1-1/w_i^\ast)$.

For a matching lower bound, we consider the special point $x^\ast\in\mathcal{X}$ such that for all $i$,
\begin{equation*}
    A^{(i)}(x^\ast,y)=\frac{1}{w_i^\ast}\pi(y)+\left(1-\frac{1}{w_i^\ast}\right)\delta_{x^\ast}(y).
\end{equation*}
Going through the full interior loop within one iteration, the whole rejection probability is at least
\begin{equation*}
    R(x^\ast)\geq\prod_{i=1}^{k}\left(1-\frac{1}{w_i^\ast}\right).
\end{equation*}
By Lemma~\ref{lem:lower}, a matching lower bound thus obtained.

\vspace{2mm}

\noindent{\bf Part 3.} We then establish \eqref{eq:generalized_rate}. For $k=2$,
	\begin{align*}
	&1-\mathbb{E}\left[\frac{1}{w_1^\ast+w_2(X_2)}\right]-\mathbb{E}\left[\frac{1}{w_1(X_1)+w_2^\ast}\right]-\left(1-\frac{1}{w_1^\ast}\right)\left(1-\frac{1}{w_2^\ast}\right)\\
	=&\mathbb{E}\left[\frac{w_1(X_1)}{w_2^\ast(w_1(X_1)+w_2^\ast)}\right]+\mathbb{E}\left[\frac{w_2(X_2)}{w_1^\ast(w_1^\ast+w_2(X_2))}\right]-\frac{1}{w_1^\ast w_2^\ast}\\
	\geq&\frac{1}{w_1^\ast(w_1^\ast+w_2^\ast)}+\frac{1}{w_2^\ast(w_1^\ast+w_2^\ast)}-\frac{1}{w_1^\ast w_2^\ast}=0.
	\end{align*}
	
	For larger $k>2$, we have, for an arbitrary fixed $l\in\{1,\ldots,k\}$,
	\begin{align}
	&1-\sum_{j=1}^{k}\mathbb{E}\left[\frac{1}{w_j^\ast+\sum_{i=1,i\neq j}^{k}w_i(X_i)}\right]\notag\\
	&-\left(1-\frac{1}{w_l^\ast}\right)\left\{1-\sum_{j=1,j\neq l}^{k}\mathbb{E}\left[\frac{1}{w_j^\ast+\sum_{i=1,i\neq j,i\neq l}^{k}w_i(X_i)}\right]\right\}\label{eqn:recursive_2}\\
	=&\sum_{j=1,j\neq l}^k \mathbb{E}\left[\frac{w_j(X_j)}{w_l^\ast[w_l^\ast+\sum_{i=1,i\neq l}^k w_i(X_i)]}\right]-\sum_{j=1,j\neq l}^k \mathbb{E}\left[\frac{1}{w_l^\ast[w_j^\ast+\sum_{i=1,i\neq j,i\neq l}^k w_i(X_i)]}\right]\notag\\
	&+\sum_{j=1,j\neq l}^k \mathbb{E}\left[\frac{w_l(X_l)}{[w_j^\ast+\sum_{i=1,i\neq j}^k w_i(X_i)][w_j^\ast+\sum_{i=1,i\neq j,i\neq l}^k w_i(X_i)]}\right]\notag\\
	\geq&\sum_{j=1,j\neq l}^k \mathbb{E}\left[\frac{1}{w_l^\ast(w_l^\ast+w_j^\ast+B_{jl})}+\frac{1}{(w_l^\ast+w_j^\ast+B_{jl})(w_j^\ast+B_{jl})}-\frac{1}{w_l^\ast(w_j^\ast+B_{jl})}\right]=0,\notag
	\end{align}
	where we denote $B_{jl}=\sum_{i=1,i\neq j,i\neq l}^k w_i(X_i)$ for simplicity. The last inequality is mainly due to
	\begin{equation*}
	    w^\ast_l+\sum_{i=1,i\neq l}^k w_i(X_i)\leq w_l^\ast+ w_j^\ast+\sum_{i=1,i\neq l,i\neq j}^k w_i(X_i)=w_l^\ast+ w_j^\ast+B_{jl}
	\end{equation*}
	applied in the denominators of the two positive terms.
	The last step of induction is the same as the proof of Theorem~\ref{comparison thm}. Suppose the result holds for $k-1$, i.e.,
	\begin{equation*}
	    1-\sum_{j=1}^{k-1}\mathbb{E}_{p}\left[\frac{1}{w_j^\ast+\sum_{1\leq i\leq k-1,i\neq j}w_i(X_i)}\right]
	\geq\prod_{i=1}^{k-1} \left(1-\frac{1}{w_i^\ast}\right),
	\end{equation*}
	it immediately follows from \eqref{eqn:recursive_2} with $l=k$ that
	\begin{align*}
	    &1-\sum_{j=1}^{k}\mathbb{E}_{p}\left[\frac{1}{w_j^\ast+\sum_{1\leq i\leq k-1,i\neq j}w_i(X_i)}\right]\\
	    \geq& \left(1-\sum_{j=1}^{k-1}\mathbb{E}_{p}\left[\frac{1}{w_j^\ast+\sum_{1\leq i\leq k-1,i\neq j}w_i(X_i)}\right]\right)\left(1-\frac{1}{w^\ast_k}\right)\\
	\geq&\prod_{i=1}^k \left(1-\frac{1}{w_i^\ast}\right).
	\end{align*}
	
	The proofs of Theorem~\ref{comparison thm} and Theorem~\ref{final ineq} are essentially the same, both utilizing induction to recursively handle a general integer $k$.
\end{proof}

\begin{proof}[Proof of Theorem~\ref{detailed balance}]
	To make our notations more explicit, we assume that every distribution mentioned here has a density with respect to the Lebesgue measure. Denote $A(x,y)$ as the actual transition density, we compute directly that
	\begin{align*}
	&\pi(x)A(x,y)\\
	=&\pi(x)\sum_{j=1}^{k}\mathbb{P}(y_j=y,J=j,y_J\text{ gets accepted})\\
	=&\pi(x)\sum_{j=1}^k \int p(x,\by_j)\frac{w_j(y,x)}{w_j(y,x)+\sum_{i\neq j}w_i(y_i,x)}\\
	&\rho p(y,\bx^\ast_{-j}\mid x) \prod_{i\neq j}\mathrm{d}y_i\mathrm{d}x_i^\ast,
	\end{align*}
	where we write $\bx^\ast_{(-j)}=(x_1^\ast,x_2^\ast,\ldots,x_{j-1}^\ast,x_{j+1}^\ast,\ldots,x_k^\ast)\in\mathcal{X}^{k-1}$ and $\by(j)=(y_1,\ldots,y_{j-1},y,y_{j+1},\ldots,y_k)\in\mathcal{X}^k$.
	Plugging in the definition of $\rho$, we use the notations $\bx^\ast(j)\triangleq(x^\ast_1,\ldots,x^\ast_{j-1},x,x^\ast_{j+1},x^\ast_k)$ and $u_j(\bx,\by)\triangleq\min\left\{\frac{1}{\sum_{i=1}^k w_i(y_i,x_j)},\frac{1}{\sum_{i=1}^kw_i(x_i,y_j)}\right\}$ to get
	\begin{align*}
	&\pi(x)A(x,y)\\
	=&\pi(x)\sum_{j=1}^k \int p(x,\by(j))\frac{w_j(y,x)}{w_j(y,x)+\sum_{i\neq j}w_i(y_i,x)}\\
	&\min\left\{1,\dfrac{w_j(y,x)+\sum_{i\neq j}w_i(y_i,x)}{w_j(x,y)+\sum_{i\neq j}w_i(x_i^\ast,y)}\right\}\\
	&p(y,\bx^\ast_{(-j)}\mid x) \prod_{i\neq j}\mathrm{d}y_i\mathrm{d}x_i^\ast\\
	=&\sum_{j=1}^k \pi(x)w_j(y,x)p_j(x,y)\int u_j(\bx^\ast(j),\by)\\
	&p(x,\by_{(-j)}\mid y_j=y)p(y,\bx^\ast_{(-j)}\mid x_j=x)\prod_{i\neq j}\mathrm{d}y_i\mathrm{d}x_i^\ast.
	\end{align*}
	In the above formula, we use the identity
	\begin{equation*}
	p(x,\by(j))=p_j(x,y)\times p(x,\by_{(-j)}\mid y_j=y).
	\end{equation*}
	At last, note that $\pi(x)w_j(y,x)p_j(x,y)=\pi(x)\pi(y)p_j(x,y)p_j(y,x)\lambda_j(x,y)$ is symmetric by our constructions, which implies that $\pi(x)A(x,y)$ is symmetric in $x$ and $y$, proving the detailed balance condition.
\end{proof}

\begin{proof}[Proof of Theorem~\ref{thm:condition}]
	If we simply set $x^\ast_j:=y_j$ for any $j\neq J$ in Algorithm~\ref{Generalized Multiple-Try Metropolis}, the conditional probability becomes
	\begin{align*}
	&\pi(x)A(x,y)\\
	=&\pi(x)\sum_{j=1}^k 
	\Bigg[\int  p(x,\by(j))\\
	&w_j(y,x)\min\left[\frac{1}{w_j(y,x)+\sum_{i\neq j}w_i(y_i,x)},\frac{1}{w_j(x,y)+\sum_{i\neq j}w_i(y_i,y)}\right]  
	\prod_{i\neq j}\mathrm{d}y_i \Bigg]\\
	=&\sum_{j=1}^k \Bigg[\int \pi(x)p_j(x,y)w_j(y,x)p(x,\by_{(-j)}\mid y_j=y) \\
	&\min\left[\frac{1}{w_j(y,x)+\sum_{i\neq j}w_i(y_i,x)},\frac{1}{w_j(x,y)+\sum_{i\neq j}w_i(y_i,y)}\right] \prod_{i\neq j}\mathrm{d}y_i \Bigg].
	\end{align*}
Since $\pi(x)w_j(y,x)p_j(x,y)$ is symmetric for $x$ and $y$, the theorem follows easily from  condition (16) in the main text.
\end{proof}




\end{appendices}

%\bibliographystyle{apalike}
\bibliography{references}% common bib file

\begin{thebibliography}{}

\bibitem[Atchad{\'e} and Perron, 2007]{atchade2007geometric}
Atchad{\'e}, Y.~F. and Perron, F. (2007).
\newblock On the geometric ergodicity of metropolis-hastings algorithms.
\newblock {\em Statistics}, 41(1):77--84.

\bibitem[B{\'e}dard et~al., 2012]{bedard2012scaling}
B{\'e}dard, M., Douc, R., and Moulines, E. (2012).
\newblock Scaling analysis of multiple-try mcmc methods.
\newblock {\em Stochastic Processes and their Applications}, 122(3):758--786.

\bibitem[Brooks et~al., 2011]{brooks2011handbook}
Brooks, S., Gelman, A., Jones, G., and Meng, X.-L. (2011).
\newblock {\em Handbook of markov chain monte carlo}.
\newblock CRC press.

\bibitem[Calderhead, 2014]{calderhead2014general}
Calderhead, B. (2014).
\newblock A general construction for parallelizing metropolis- hastings
  algorithms.
\newblock {\em Proceedings of the National Academy of Sciences},
  111(49):17408--17413.

\bibitem[Casarin et~al., 2013]{casarin2013interacting}
Casarin, R., Craiu, R., and Leisen, F. (2013).
\newblock Interacting multiple try algorithms with different proposal
  distributions.
\newblock {\em Statistics and Computing}, 23(2):185--200.

\bibitem[Chen et~al., 1994]{chen1994weighted}
Chen, X.-H., Dempster, A.~P., and Liu, J.~S. (1994).
\newblock Weighted finite population sampling to maximize entropy.
\newblock {\em Biometrika}, 81(3):457--469.

\bibitem[Craiu and Lemieux, 2007]{craiu2007acceleration}
Craiu, R.~V. and Lemieux, C. (2007).
\newblock Acceleration of the multiple-try metropolis algorithm using
  antithetic and stratified sampling.
\newblock {\em Statistics and computing}, 17(2):109--120.

\bibitem[Dai and Liu, 2020]{dai2020monte}
Dai, C. and Liu, J.~S. (2020).
\newblock Monte carlo approximation of bayes factors via mixing with surrogate
  distributions.
\newblock {\em Journal of the American Statistical Association}, pages 1--16.

\bibitem[Diaconis et~al., 2008]{diaconis2008gibbs}
Diaconis, P., Khare, K., and Saloff-Coste, L. (2008).
\newblock Gibbs sampling, exponential families and orthogonal polynomials.
\newblock {\em Statistical Science}, 23(2):151--178.

\bibitem[Diaconis and Saloff-Coste, 1998]{diaconis1998we}
Diaconis, P. and Saloff-Coste, L. (1998).
\newblock What do we know about the metropolis algorithm?
\newblock {\em Journal of Computer and System Sciences}, 57(1):20--36.

\bibitem[Frenkel et~al., 1996]{frenkel1996understanding}
Frenkel, D., Smit, B., and Ratner, M.~A. (1996).
\newblock {\em Understanding molecular simulation: from algorithms to
  applications}.
\newblock Academic press San Diego.

\bibitem[Hastings, 1970]{hastings1970monte}
Hastings, W.~K. (1970).
\newblock Monte carlo sampling methods using markov chains and their
  applications.

\bibitem[Levin and Peres, 2017]{levin2017markov}
Levin, D.~A. and Peres, Y. (2017).
\newblock {\em Markov chains and mixing times}, volume 107.
\newblock American Mathematical Soc.

\bibitem[Liu, 1996]{liu1996metropolized}
Liu, J.~S. (1996).
\newblock Metropolized independent sampling with comparisons to rejection
  sampling and importance sampling.
\newblock {\em Statistics and computing}, 6(2):113--119.

\bibitem[Liu, 2008]{liu2008monte}
Liu, J.~S. (2008).
\newblock {\em Monte Carlo strategies in scientific computing}.
\newblock Springer Science \& Business Media.

\bibitem[Liu et~al., 2000]{liu2000multiple}
Liu, J.~S., Liang, F., and Wong, W.~H. (2000).
\newblock The multiple-try method and local optimization in metropolis
  sampling.
\newblock {\em Journal of the American Statistical Association},
  95(449):121--134.

\bibitem[Liu et~al., 1995]{liu1995covariance}
Liu, J.~S., Wong, W.~H., and Kong, A. (1995).
\newblock Covariance structure and convergence rate of the gibbs sampler with
  various scans.
\newblock {\em Journal of the Royal Statistical Society: Series B
  (Methodological)}, 57(1):157--169.

\bibitem[Martino, 2018]{martino2018review}
Martino, L. (2018).
\newblock A review of multiple try mcmc algorithms for signal processing.
\newblock {\em Digital Signal Processing}, 75:134--152.

\bibitem[Martino et~al., 2014]{martino2014multiple}
Martino, L., Leisen, F., and Corander, J. (2014).
\newblock On multiple try schemes and the particle metropolis-hastings
  algorithm.
\newblock {\em arXiv preprint arXiv:1409.0051}.

\bibitem[Metropolis et~al., 1953]{metropolis1953equation}
Metropolis, N., Rosenbluth, A.~W., Rosenbluth, M.~N., Teller, A.~H., and
  Teller, E. (1953).
\newblock Equation of state calculations by fast computing machines.
\newblock {\em The journal of chemical physics}, 21(6):1087--1092.

\bibitem[Neal, 2011]{neal2011mcmc}
Neal, R.~M. (2011).
\newblock Mcmc using ensembles of states for problems with fast and slow
  variables such as gaussian process regression.
\newblock {\em arXiv preprint arXiv:1101.0387}.

\bibitem[Pandolfi et~al., 2010]{pandolfi2010generalization}
Pandolfi, S., Bartolucci, F., and Friel, N. (2010).
\newblock A generalization of the multiple-try metropolis algorithm for
  bayesian estimation and model selection.
\newblock In {\em Proceedings of the thirteenth international conference on
  artificial intelligence and statistics}, pages 581--588. JMLR Workshop and
  Conference Proceedings.

\bibitem[Qin and Liu, 2001]{qin2001multipoint}
Qin, Z.~S. and Liu, J.~S. (2001).
\newblock Multipoint metropolis method with application to hybrid monte carlo.
\newblock {\em Journal of Computational Physics}, 172(2):827--840.

\bibitem[Roberts and Tweedie, 1996]{roberts1996geometric}
Roberts, G.~O. and Tweedie, R.~L. (1996).
\newblock Geometric convergence and central limit theorems for multidimensional
  hastings and metropolis algorithms.
\newblock {\em Biometrika}, 83(1):95--110.

\bibitem[Tierney, 1994]{tierney1994markov}
Tierney, L. (1994).
\newblock Markov chains for exploring posterior distributions.
\newblock {\em the Annals of Statistics}, pages 1701--1728.

\bibitem[Wang, 2020]{wang2020exact}
Wang, G. (2020).
\newblock Exact convergence analysis of the independent metropolis-hastings
  algorithms.
\newblock {\em arXiv preprint arXiv:2008.02455}.

\bibitem[Yang et~al., 2018]{yang2018parallelizable}
Yang, S., Chen, Y., Bernton, E., and Liu, J.~S. (2018).
\newblock On parallelizable markov chain monte carlo algorithms with
  waste-recycling.
\newblock {\em Statistics and Computing}, 28(5):1073--1081.

\end{thebibliography}
%\bibliographystyle{apalike}
%\input sn-mathphys.bst

\end{document}